\documentclass[iop]{emulateapj}
\usepackage{epsfig,amsmath}
\usepackage[dvips]{color}
\usepackage{verbatim}
\usepackage{graphicx}
\usepackage{txfonts}
\usepackage{natbib}
\usepackage{hyperref}
\usepackage{color}

\newcommand{\IUE}{{\it IUE}}          
\newcommand{\HST}{{\it HST}}
\newcommand{\kms}{\ifmmode {\rm km\ s}^{-1} \else km s$^{-1}$\fi}
\newcommand{\Msun}{\ifmmode {\rm M}_{\odot} \else M$_{\odot}$\fi}
\newcommand{\Lsun}{\ifmmode {\rm L}_{\odot} \else L$_{\odot}$\fi}
\newcommand{\qo}{\ifmmode q_{\rm o} \else $q_{\rm o}$\fi}
\newcommand{\Ho}{\ifmmode H_{\rm o} \else $H_{\rm o}$\fi}
\newcommand{\ho}{\ifmmode h_{\rm o} \else $h_{\rm o}$\fi}
\newcommand{\ltsim}{\raisebox{-.5ex}{$\;\stackrel{<}{\sim}\;$}}

\newcommand{\vFWHM}{\ifmmode v_{\mbox{\tiny FWHM}} \else
                    $v_{\mbox{\tiny FWHM}}$\fi}
\newcommand{\CCF}{\ifmmode F_{\it CCF} \else $F_{\it CCF}$\fi}
\newcommand{\ACF}{\ifmmode F_{\it ACF} \else $F_{\it ACF}$\fi}
\newcommand{\Halpha}{\ifmmode {\rm H}\alpha \else H$\alpha$\fi}
\newcommand{\Hbeta}{\ifmmode {\rm H}\beta \else H$\beta$\fi}
\newcommand{\hbeta}{\ifmmode {\rm H}\beta \else H$\beta$\fi}
\newcommand{\Hgamma}{\ifmmode {\rm H}\gamma \else H$\gamma$\fi}
\newcommand{\Hdelta}{\ifmmode {\rm H}\delta \else H$\delta$\fi}
\newcommand{\Lya}{\ifmmode {\rm Ly}\alpha \else Ly$\alpha$\fi}
\newcommand{\Lyb}{\ifmmode {\rm Ly}\beta \else Ly$\beta$\fi}
\newcommand{\HeI}{\ifmmode {\rm He}\,{\sc i}\,\lambda5876 \else 
	          He\,{\sc i}\,$\lambda5876$\fi}
\newcommand{\HeII}{\ifmmode {\rm He}\,{\sc ii}\,\lambda4686 \else 
	           He\,{\sc ii}\,$\lambda4686$\fi}

\newcommand{\heii}{He\,{\sc ii}}

\newcommand{\feii}{Fe\,{\sc ii}}

\newcommand{\ciii}{\ifmmode {\rm C}\,{\sc iii} \else C\,{\sc iii}\fi}
\newcommand{\civ}{\ifmmode {\rm C}\,{\sc iv} \else C\,{\sc iv}\fi}

\newcommand{\oiii}{O\,{\sc iii}}

\newcommand{\Flamunit}{\ifmmode {\rm erg\ s}^{-1} cm^{-2}\ \AA^{-1} \else erg s$^{-1}$\ cm$^{-2}$\,\AA$^{-1}$\fi}

\newcommand{\lboli}{\ifmmode L^{ion}_{BOL} \else $L^{ion}_{BOL}$\fi}

\shortauthors{Kilerci, Vestergaard, Peterson, Denney, Bentz}
\shorttitle{On Scatter in the AGN Radius $-$  Luminosity Relationship}

\accepted{9. November 2014}

\begin{document}
  
\title{ON THE SCATTER IN THE RADIUS  $-$  LUMINOSITY RELATIONSHIP FOR ACTIVE GALACTIC NUCLEI}

\author{ E.~Kilerci Eser\altaffilmark{1},
            M.~Vestergaard\altaffilmark{1,2},
            B.~M.~Peterson\altaffilmark{3,4},
            K.~D.~Denney\altaffilmark{3,5},
            M.~C.~Bentz\altaffilmark{6}
          }
\altaffiltext{1}{Dark Cosmology Centre, Niels Bohr Institute, University of Copenhagen,
Juliane Maries Vej 30, DK-2100 Copenhagen \O, Denmark; ecekilerci, vester @dark-cosmology.dk}
\altaffiltext{2}{Steward Observatory, 
		University of Arizona, 
		933 North Cherry Avenue, 
		Tucson, AZ 85721.}
\altaffiltext{3}{Department of Astronomy, 
		The Ohio State University, 
		140 West 18th Avenue, 
		Columbus, OH 43210; kelly, peterson@astronomy.ohio-state.edu}
\altaffiltext{4}{The Center for Cosmology and AstroParticle Physics, 
		The Ohio State University, 
		191 West Woodruff Avenue, Columbus. OH 43210}
\altaffiltext{5}{National Science Foundation Fellow}
\altaffiltext{6}{Department of Physics and Astronomy, 
		Georgia State University, Atlanta, GA 30303; bentz@chara.gsu.edu}

\begin{abstract}
We investigate and quantify the observed scatter in the empirical relationship between the broad line 
region size $R$ and the luminosity of the active galactic nucleus (AGN), in order to better understand its origin.
This study is motivated by 
the indispensable role of this relationship in the mass estimation of cosmologically distant black holes, but may also be relevant to the recently proposed application of this relationship for measuring cosmic distances.
We study six nearby reverberation-mapped AGN for which simultaneous UV and optical monitoring 
data exist. 
We also examine the long-term optical luminosity variations of Seyfert 1 galaxy NGC\,5548 and employ Monte Carlo simulations to study the effects of the intrinsic variability of individual objects on the scatter in the global relationship for a sample of $\sim$40 AGN. 
We find the scatter in this relationship has a correctable dependence on color. 
For individual AGN, the size of the $\Hbeta$ emitting region has a steeper 
dependence on the nuclear optical luminosity than on the UV luminosity, that can introduce a scatter of $\sim$0.08\,dex into the global relationship,
due the non-linear relationship between the variations in the ionizing continuum and those in the optical continuum. 
Also, our analysis highlights the importance of understanding and minimizing the scatter in the relationship traced by the intrinsic variability of individual AGN, since it propagates directly into the global relationship. 
We find that using the UV luminosity as a substitute for the ionizing luminosity can reduce a sizable fraction of the 
current observed scatter of $\sim$0.13\,dex. 

\end{abstract}



\section{INTRODUCTION\label{introduction}}

Owing to their powerful and persistent emission that can be observed across most of the observable Universe
\citep[e.g.,][]{Mortlock2011},  there has been a strong interest in using quasars 
as cosmological probes since their discovery.  Because active galactic nuclei (AGN) and quasars are powered by accretion of matter
onto supermassive black holes \citep{Lynden-Bell1969, Rees1984} centered in their host galaxies, and the majority reside at cosmic distances
\citep[e.g.,][]{WHO1994, Fan2001a},  there are multiple ways in which these enigmatic sources can be used as cosmic probes: 
\begin{enumerate}
\item 
A quasar can be used as a background light source to study the intervening intergalactic medium as it absorbs the quasar emission \citep[e.g.,][]{Wolfe2005,Krogager2013,Fynbo2013}; 
\item 
Quasars can act as `light houses' by which to locate and study some of the most massive galaxies in the Universe out to the earliest epochs. This can be done because quasars are powered by the most massive black holes known 
\citep[e.g.,][]{Vestergaard2004, Vestergaard2008, Jiang2010,DeRosa2014}, 
and the most massive black holes tend to reside in the most massive galaxies \citep[e.g.,][]{Tremaine2002, FerrareseFord2005}; 
\item  
Beyond the local Universe, direct measurements of the mass of the central black hole is not possible for quiescent galaxies \citep[e.g.,][]{FerrareseFord2005}. Only for AGN and quasars can the black hole mass be measured in this case, 
permitting studies of black hole growth \citep[e.g.,][]{Vestergaard2009, Kelly2010, Trakhtenbroot2012, KellyShen2013} and feedback $-$ as manifested in observations of galaxy clusters \citep[e.g.,][]{McNamaraNulsen2007} and galaxies \citep[e.g.,][]{Merloni2007,Werner2013} and in numerical simulations \citep[e.g.,][]{Hopkins2006, Croton2006}; 
\item 
An AGN or quasar can act as a standard candle or standard ruler to measure cosmological distances or constrain cosmological parameters \citep[e.g.,][]{Collier1999, Elvis2002, Cackett2007, Watson2011, Haas2011, King2014, Honig2014, Yoshii2014}.  
\end{enumerate}

The use of AGN as standard candles/rulers has previously been attempted by means of the broad line equivalent width \citep[i.e., the Baldwin Effect;][]{Baldwin1977} and/or accretion disk emission \citep[e.g.,][]{Collier1999, Elvis2002, Cackett2007}, but neither method has yet proven particularly useful.
The situation has changed in the last few years as the empirical relationship between the `size' (or radius, $R$) of the broad emission line region (BLR) and the nuclear continuum luminosity $L$ \citep[i.e., the $R$\,$-$\,$L$  relationship; e.g.,][and references therein]{Kaspi2000, Bentz2013} has proven to be especially tight, permitting a more robust measure of the AGN luminosity.  While the relationship has traditionally been used to predict the BLR distance from the black hole for estimates of black hole masses of distant quasars \citep[e.g.,][]{Vestergaard2002, McLureJarvis2002, Vestergaard2006, McGill2008, Wang2009, Rafiee2011, Shen2011, Park2013},
recent studies suggest its use as a cosmological probe also at high redshifts \citep{Watson2011, Haas2011, Melia2014}. 
In particular, \citet{Watson2011} suggest the reverse use of the relationship to predict the luminosity from a direct measure of the BLR size and propose ways in which the scatter in the relationship at the time ($\sim$0.2\,dex, corresponding to a distance modulus $\Delta \mu$=0.5\,mag) can be reduced.

\paragraph{The AGN Radius $-$ Luminosity Relationship}

The emission from the central engine in AGN and quasars is not constant in time, but varies, likely in response to variations in the rate at which matter falls onto the supermassive black hole from the accretion disk surrounding it. Gas in their immediate vicinity, the so-called broad line region, is photoionized by the continuum photons emitted by the central accretion disk and emits the characteristic broad emission lines that are among the defining spectral features of Type 1 AGN.  
The emission line fluxes vary in response to the changes in the driving continuum luminosity with a certain time delay, $\tau$. This delay is the light travel time of the ionizing photons to the BLR, and we can infer the size of this region, i.e., the distance to the gas, as: $R_{BLR}=c \tau$, where $c$ is the light speed. The reverberation-mapping (RM) technique \citep{Blandford82, Peterson93} measures $\tau$ by comparing the continuum and line emission light curves. There are now nearly 50 measurements of the size of the $\Hbeta$ broad line-emitting regions, $R(\Hbeta)$, in nearby AGN \citep[][and references therein]{Peterson04,Bentz2010,Bentz2013} plus several measurements of lags for other emission lines. 
We observe a relatively tight relationship between the size $R_{BLR}$ and the optical nuclear continuum luminosity, $L$(optical) \citep[e.g.] [and references therein]{Bentz09a, Bentz2013}.  
In the following, we use $R_{BLR}$ to refer to the BLR size in general, and $R(\Hbeta)$ and $R$(\civ) to refer to the sizes of the $\Hbeta$ and \civ{} emitting regions, respectively.

The empirically established $R_{BLR} - L$ relationship is expected from the underlying 
photoionization physics \citep[e.g.,][]{Peterson2002, Korista2004}.  The main parameters of photoionization equilibrium models are: 
(i) elemental abundances, (ii) the shape of the ionizing continuum, (iii) 
the particle density of the photo-ionized gas, and (iv) the ionization parameter 
$U$ \citep{Osterbrock2006} defined for hydrogen as: 
\begin{equation}\label{E:ion} 
U=\frac{1}{4\pi R^{2}cn_{H}}\int\limits_{\nu_0}\limits^{\infty}\frac{L_{\nu}}{h\nu}d\nu =\frac{Q(H)}{4\pi R^{2}cn_{H}} \propto\frac{L({\rm ionizing})}{4\pi R^{2}cn_{H}}
\end{equation} 
where $n_{H}$ is the total hydrogen number density; $R$ is the distance to the ionized gas 
(here, it is the BLR radius for the hydrogen broad emission lines); ${\nu_0}$ is the threshold ionization frequency for hydrogen;  
$Q(H)$ is the production rate of hydrogen ionizing photons and $L$(ionizing) is the ionizing luminosity. 
To first order, AGN spectra look the same across a wide range in luminosity \citep{DavidsonNetzer79,Baldwin1995,Dietrich2002}. 
This suggests that the values of $n_H$ and $U$ (or the product $U n_H$)
are generally the same for all BLRs 
\citep[e.g.,][]{Petersonbook}. Under this assumption, the distance to the line emitting gas is 
expected to scale as $R_{BLR}\propto L({\rm ionizing})^{0.5}$. 

There have been several attempts in the past 20 years to test the existence of the 
$R_{BLR}-L$ relationship and to measure its slope.  
\citet{Davidson1972} was the first to emphasize the 
importance of the ionization parameter in early photoionization calculations.
The $R_{BLR}-L$ relationship appeared explicitly in the early 
reviews that covered emission-line variability \citep{Mathews1984,Peterson1988}. 
The first attempts at establishing the relationship were made in the early 1990's
\citep[e.g.][]{Koratkar1991,Peterson93} based on the early compilations of the first reverberation data.  \citet{Laor1998} and \citet*{Wandel1999} used 
the reverberation data available at the time for the first calibration of the black hole 
mass scale based on radii calculated from the photoionization formula. 
The observed $R_{BLR}-L(5100\AA)$ relationship finally became convincing with the addition of higher luminosity quasars \citep{Kaspi2000} that not only doubled the size of the reverberation database 
but also expanded the luminosity range by another two orders of magnitude.

Although the larger reverberation mapping sample size solidified the existence of an $R_{BLR}-L$(5100\AA) relationship, the observed slope \citep{Kaspi2000, Kaspi2005} was steeper than that expected from photoionization physics $-$ a consequence, it turns out, of reverberation mapping campaign observing strategies.
The large aperture used for accurate spectrophotometry lets in more host galaxy light and the 
observed continuum luminosity, therefore, contains an unwanted contribution from star light that can 
be significant for nearby AGN and is relatively larger for Seyferts than for quasars. 
This is so for two reasons: (1) Seyferts tend to be nearby objects for which the host galaxies are larger 
and brighter on the sky, and (2) the large intrinsic brightness of the nuclear source 
in quasars results in a large contrast of this emission relative to that of its host galaxy.  
Using $\HST$ and ground-based imaging \citet{Bentz2006a,Bentz09a,Bentz2013} 
determine the host star light contribution to $L(5100\AA)$ for the reverberation-mapped 
AGN sample \citep{Peterson04, Bentz09b, Denney10, Grier2012b}. 
Based on the most recent corrected AGN luminosities that also account for the recently 
updated extinction maps for our Galaxy \citep{Schlafly2011}, \citet{Bentz2013} present 
the most well determined $R(\Hbeta)-L(5100\AA)$ relationship for the $\Hbeta$ line 
emission to date and measure a slope of 0.53, consistent with the theoretical prediction of 
the slope of 0.5 to within the errors ($\pm\sigma=0.03$\,dex). 

\begin{deluxetable*}{ccccccc}
\tablecolumns{7}
\tabletypesize{\footnotesize}
\tablewidth{0pt}
\tablecaption{Reverberation Mapping Sub-sample}
\tablehead{
\colhead{Object} &
\colhead{Redshift\tablenotemark{a}} &
\colhead{$E(B-V)$\tablenotemark{a}} &
\colhead{Distance\tablenotemark{b}}&
\colhead{Host Flux\tablenotemark{c}}&
\colhead{Aperture}&
\colhead{P.A} \\
\colhead{} &
\colhead{} &
\colhead{} &
\colhead{(Mpc)} &
\colhead{($10^{-15}$ erg s$^{-1}$ cm$^{-2}$ \AA$^{-1}$)}&
\colhead{$(\arcsec \times \arcsec)$} &
\colhead{$(^{\circ})$} \\
\colhead{(1)} &
\colhead{(2)} &
\colhead{(3)} &
\colhead{(4)} &
\colhead{(5)} &
\colhead{(6)} &
\colhead{(7)}}
\startdata
Fairall 9     & 0.04702 & 0.023 &202.8$\pm$7.2 &3.21 $\pm$ 0.16& $4.0 \times9.0$ &0.0\\ 
3C 390.3   & 0.05610 & 0.063 &243.5$\pm$7.2 &0.99 $\pm$ 0.05&$5.0 \times7.5$  &90.0\\ 
NGC 7469& 0.01632 & 0.061 &68.8$\pm$7.0  &10.18 $\pm$ 0.94 &$5.0 \times7.5$  &90.0\\ 
NGC 5548& 0.01718 & 0.018 &72.5$\pm$7.0  &3.97 $\pm$ 0.40 &$5.0 \times7.5$  &90.0\\ 
NGC 3783& 0.00973 &0.105  &25.1$\pm$5.0  &6.55 $\pm$ 0.65 &$5.0 \times10.0$&0.0\\  
NGC 4151& 0.00332 & 0.024 &16.6$\pm$3.3  &16.17 $\pm$1.51&$5.0 \times7.5$  &90.0\\  
\enddata
\tablenotetext{a}{Redshifts and $E(B-V)$  
values are adopted from NASA/IPAC Extragalactic Database. $E(B-V)$ 
values are based on the \citet{Schlafly2011} dust maps.}
\tablenotetext{b}{\,Luminosity distances calculated from the redshifts with exception of NGC\,3783 and NGC\,4151, for which we adopt the the distances determined by \citet{Tully2009}.
They more reliable than the redshift-based 
distance because these two AGN have large peculiar velocities relative to the Hubble flow.
The distances and the associated uncertainties for NGC\,3783 and NGC\,4151 are adopted from \citet{Bentz2013},
while we assign an uncertainty of 500\,km\,s$^{-1}$ in recession velocity for the remaining distance uncertainties.}
\tablenotetext{c}{ Host galaxy flux densities, contaminating the spectral data, 
are adopted from \citet{Bentz09a,Bentz2013} and corrected for Galactic reddening 
as described in \S \ref{S:2.1.2}. 
For NGC\,4151, the host galaxy flux is a new measurement for the specified 
spectroscopic aperture.}
\label{tab:objects}
\end{deluxetable*}

The slope of the `global'  $R_{BLR}-L$(5100\AA) relationship (i.e., that traced by a sample of AGN with different black hole mass and intrinsic accretion state) is consistent with expectations based on photoionization physics, because the optical and ionizing luminosities are related (see also \S~\ref{S:dis3}).
However, $L(5100\AA)$ is only a proxy for the ionizing luminosity that drives the changes in $R_{BLR}$.
We cannot directly observe or measure $L$(ionizing) ($\lambda < 912$\AA) due to absorption by Galactic hydrogen.
\citet{Bentz2007} found that on the scale of an individual AGN, that of NGC\,5548, the 
single-object (or `native') $R_{\Hbeta}-L(5100\AA)$ relationship (i.e., that traced by its intrinsic variability and formed from multiple RM campaigns of this object)
has a slope of $\sim$0.7 that is statistically different from the photoionization physics expectations.
Yet, \citeauthor{Bentz2007} examine the empirical relationship between simultaneous pairs of optical and UV flux measurements and, combined with the available H$\beta$ lags at the time, estimate a slope of 0.55 for the native $R(\Hbeta)-L(UV)$ relationship for NGC\,5548. 
These results indicate the likelihood not only that 
$L(UV)$ is a better proxy for $L$(ionizing) than $L(5100\AA)$, but also
that the movement of individual objects along their own native
$R_{BLR}-L(5100\AA)$ relationships, as they vary, is a source of scatter
in the global $R_{BLR}-L(5100\AA)$ relationship.
And because AGN and quasars are known to become bluer when brighter, i.e., the UV variability amplitudes exceed those of the optical emission \citep[e.g.,][]{Clavel91, Kinney1991, Paltani1994, Korista95, Vandenberk2004, Wilhite2005, Meusinger2011, Zuo2012}, this is expected to impact the $R_{BLR}-L$(5100\AA) and $R_{BLR}-L(UV)$ relationships differently, perhaps through different slopes and/or scatter. 

Motivated by the growing interest to investigate possible ways to improve the methods by which quasars and AGN can be used as cosmic probes, we examine in this work the scatter in the AGN $R$\,$-$\,$L$ relationship, since it is the heart of quasar black hole mass estimates and of the quasar distance indicator method. 
In particular, we are interested in the amount of scatter that may be
attributed to the global $R_{BLR}-L$(5100\AA) relationship by the use of
$L$(5100\AA) as a stand-in for $L$(ionizing), and whether such scatter
can be mitigated by adopting a better proxy.  In the following, \S\,\ref{S:S2}
describes the sample and database used for our analyses presented in
\S3.  In \S~\ref{S:3.1} we examine the $L$(optical)$-L$(UV) relationship
for a small sample of nearby AGN for which near-simultaneous UV and
optical luminosity observations exist.  In \S\,\ref{S:3.3} and Appendix\,\ref{S:montecarlo}, we
investigate the effect that the steep native $R_{\Hbeta}-L$(5100\AA)
relationship of Seyfert 1 galaxy NGC\,5548 \citep[slope
$\sim$0.7;][]{Bentz2007,Zu2011} has on the scatter in the global
$R_{BLR}-L$(5100\AA) relationship and consider the extension of such an
effect for the larger RM sample in the global relationship.  We
discuss our results in \S4 and summarize our conclusions in \S5. A
cosmology with $H_0=72$\,km\,s$^{-1}$\,Mpc$^{-1}$, $\Omega_\Lambda =
0.7$ and $\Omega_{\rm m}=0.3$ is adopted throughout.
\newpage

\begin{deluxetable*}{ccccccc}
\tablecolumns{7}
\tabletypesize{\footnotesize}
\tablewidth{0pt}
\tablecaption{Datasets and References}
\tablehead{
\colhead{Object} &
\colhead{Optical Data} &
\colhead{Optical Data\tablenotemark{a}} &
\colhead{UV Data}&
\colhead{UV Data\tablenotemark{a}} &
\colhead{$\lambda_{rest}$} &  
\colhead{UV Continuum} \\
\colhead{} &
\colhead{Julian Dates} &
\colhead{References} &
\colhead{Julian Dates} &
\colhead{References}&
\colhead{for $\lambda L_{\lambda}(\rm UV)$ } &
\colhead{Window Range}\\
\colhead{(1)} &
\colhead{(2)} &
\colhead{(3)} &
\colhead{(4)} &
\colhead{(5)} &
\colhead{(6)} &
\colhead{(7)}}
\startdata
Fairall 9&$2449476-2449664$ &1&$2449477-2449665$&8 & 1327\AA & 30\AA\\ 
3C 390.3 &$2449734-2450068$ &2&$2449735-2450068$&9& 1297\AA & 50\AA\\
NGC 7469&$2450249-2450274$ & 3\tablenotemark{b}&$2450248-2450273$&10& 1294\AA & 20\AA\\
NGC 5548&$2447509-2447746$ &4&$2447510-2447745$&10& 1350\AA& 40\AA\\
NGC 5548 &$2449095-2449133$ &4&$2449097-2449135$&11& 1350\AA &10\AA\\
NGC 3783&$2448610-2448832$  &5&$2448612-2448833 $&12& 1445\AA & 30\AA\\   
NGC 4151&$2449318-2449335$  &6\tablenotemark{c}&$2449318-2449335$&13& 1275\AA & 30\AA\\  
\enddata
\tablenotetext{a}{References: (1) \citet{Santos97}; (2) \citet{Dietrich98}; 
(3) \citet{Collier98}; (4) \citet{Peterson13};  
(5) \citet{Stirpe94};  (6) \citet{Kaspi96};  
(7) \citet{Rodriguez-Pascual97}; 
(8) \citet{O'Brien98}; (9) \citet{ Wanders97}; (10) \citet{Clavel91}; 
(11) \citet{Korista95}; (12) \citet{Reichert94}; (13) \citet{Crenshaw96}.} 
\tablenotetext{b}{ For the NGC7469 spectra described by \citet{Collier98} we chose only 
those obtained through the $5"\times 7.5"$ aperture for which we have host galaxy flux 
density measurements \citep{Bentz09a,Bentz2013}. 
The observed flux densities at rest frame 5100\AA{} 
that we re-measured for this work are listed in Table\,\ref{tab:Fcont}.}
\tablenotetext{c}{ The NGC4151 data from \citet{Kaspi96} are the subset obtained 
with the Perkins 1.8-m telescope at Lowell Observatory with a $5"\times 7.5"$ 
spectroscopic aperture, re-calibrated for this study using the [\oiii]\ flux from 
\citet{Bentz2006}.}
\label{tab:dataset}
\end{deluxetable*}

\section{The Sample and Data}\label{S:S2}

\subsection{The Database for the Optical-UV Luminosity Relationship}\label{S:2.1}
We select six sources (NGC\,5548, NGC\,7469, NGC\,3783, NGC\,4151, 3C\,390.3, and Fairall\,9) 
from the sample of reverberation-mapped nearby AGN \citep{Peterson04} based on the 
availability of multiple epochs of quasi-simultaneous optical and UV data. 
Some basic properties of these objects (hereafter referred to as `the RM sub-sample') 
are listed in Table\,\ref{tab:objects}. 
Our study is based on the publicly available optical and UV spectroscopic data from the 
International AGN Watch\footnote{http://www.astronomy.ohio-state.edu/$\sim$agnwatch/} 
database. The UV luminosities are derived from \IUE\ and \HST\ spectral data. Accurate host galaxy fluxes \citep{Bentz2013} are available for all objects in this study. 
Each optical flux density measurement is matched with a single-epoch UV flux density that
is the temporally closest UV luminosity measurement to within two days.  When there is more 
than one observation in one day, we adopt the mean flux density and consider this daily average as `one epoch'.

We compute the rest frame monochromatic luminosity as 
$L_{\lambda ({\rm rest})}= (1+z) F_{\lambda(obs)} 4\pi D_{L}^2$, where 
$F_{\lambda(obs)}$ is the observed monochromatic flux density, 
$z$ is the redshift, $\lambda $(obs) = $\lambda $(rest)$(1+z)$, and  $D_{L}$ is the 
luminosity distance of the source; the values of $z$ and $D_{L}$ adopted here are 
listed in Table\,\ref{tab:objects}. For NGC\,4151 and NGC\,3783 we adopt the 
distances determined by \citet{Tully2009} because these galaxies are so nearby that the Hubble flow 
distance is inaccurate.

The optical monochromatic flux density is the average flux density in 
a $\sim$20\,\AA{} $-$ 30\, \AA{} wide range centered at a rest frame wavelength of 5100\AA. 
For 3C\,390.3, the optical flux density is measured between $5170\AA$ and $5180\AA$
since the wavelength region around 5100\AA\ is contaminated by \feii\ emission \citep{Dietrich98}.
Similarly, the UV continuum fluxes are the mean flux densities measured over a range 
of $\sim$30\,\AA{} in the UV spectra.
Table~\ref{tab:dataset} presents the object names (column 1), the Julian dates of the 
optical and UV observations (columns 2 and 4, respectively), and the wavelength range (column 7)
centered at the specific rest frame wavelength (column 6) over which 
the monochromatic UV luminosities were measured. 
  References to the original studies that first presented these data are listed in columns 
3 and 5. In the following, we address the calibrations and the corrections applied to 
the data prior to our analyses.

\subsubsection{Calibration and Measurements of the RM Sub-sample Data }\label{S:2.1.1}

In reverberation-mapping studies, it is common to use the [\oiii]$\lambda\lambda 4959, 5007$ line 
emission as an internal flux calibrator to place the spectra on an absolute flux scale. 
Internal flux calibration is necessary to account for varying atmospheric transparency, 
seeing conditions and potential  slit losses due to seeing changes during the observations. 
The internal flux calibration is based on the assumption that the [\oiii]\ line emission 
is constant over the variability time scale ($\sim$days to weeks) of the broad line emission.
This is a reasonable assumption because the [\oiii]\ line flux is typically constant 
on timescales of many years \citep{Peterson93} because it is produced by the narrow 
line gas located at spatial scales of $\sim$100 pc, much farther from the BLR, and 
because the narrow-line region gas density is so low, that the recombination time 
scale is also very long. 
All the data analyzed here are calibrated by scaling the observed 
[\oiii]$\lambda 5007$ line flux to an absolute [\oiii]\ flux measurement 
based on spectrophotometric observations (see Table\,\ref{tab:dataset} for references). 
Correction of the calibrated flux densities for reddening and host galaxy contribution
is addressed in \S\ref{S:2.1.2} and \S\ref{S:2.1.3}, respectively.

\begin{deluxetable*}{lcccc} 
\tablecolumns{5}
\tablewidth{0pt}
\tablecaption{Updated NGC\,5548 Mean Flux Densities and Luminosities}
\tablehead{
\colhead{Data Set} &
\colhead{$F_{var}(continuum)$\tablenotemark{a}} &
\colhead{{F(5100\AA)\tablenotemark{a}}$\pm\sigma$\tablenotemark{b}} &
\colhead{$\log [\lambda L_{\lambda}(5100\AA) /erg\,s^{-1}]$\tablenotemark{c}}  &
\colhead{$\log [L(\Hbeta) /erg\,s^{-1}]$\tablenotemark{a,d}}  \\
\colhead{} &
\colhead{} &
\colhead{$(10^{-15} erg\,s^{-1} cm^{-2} \AA^{-1})$}&
\colhead{$\pm \sigma$}&
\colhead{$\pm \sigma$}\\
\colhead{(1)} & \colhead{(2)} & \colhead{(3)} & \colhead{(4)} & \colhead{(5)}}
\startdata
Year 1  &  0.188&6.176 $\pm$ 0.648  &43.33  $\pm$   0.09 & 41.70 $\pm$ 0.04 \\ 
Year 2  &  0.272&3.378 $\pm$ 0.546  &43.07  $\pm$   0.10 & 41.55 $\pm$ 0.05 \\ 
Year 3  &  0.154&5.336 $\pm$ 0.551  &43.26  $\pm$   0.09 & 41.63 $\pm$ 0.06 \\ 
Year 4  &  0.386&2.901 $\pm$ 0.504  &43.00  $\pm$   0.10 & 41.42 $\pm$ 0.05 \\ 
Year 5  &  0.148&5.375 $\pm$ 0.546  &43.27  $\pm$   0.09 & 41.67 $\pm$ 0.06 \\ 
Year 6  &  0.173&5.620 $\pm$ 0.588  &43.29  $\pm$   0.09 & 41.63 $\pm$ 0.04 \\
Year 7  &  0.117&7.918 $\pm$ 0.508  &43.44  $\pm$   0.08 & 41.74 $\pm$ 0.04 \\
Year 8  &  0.244&6.021 $\pm$ 0.554  &43.32  $\pm$   0.08 & 41.67 $\pm$ 0.04 \\ 
Year 9  &  0.209&3.765 $\pm$ 0.509  &43.11  $\pm$   0.09 & 41.61 $\pm$ 0.09 \\ 
Year 10  &  0.146&8.344 $\pm$ 0.630  &43.46  $\pm$   0.08 & 41.80 $\pm$ 0.03 \\ 
Year 11  &  0.229&6.899 $\pm$ 0.597  &43.38  $\pm$   0.08 & 41.70 $\pm$ 0.06 \\ 
Year 12  &  0.424&2.407 $\pm$ 0.502  &42.92  $\pm$   0.11 & 41.51 $\pm$ 0.04 \\ 
Year 13  &  0.293&2.323 $\pm$ 0.510  &42.90  $\pm$   0.11 & 41.40 $\pm$ 0.05 \\ 
Year 17  &  0.187&0.975 $\pm$ 0.527  &42.53  $\pm$   0.20 & 41.01 $\pm$ 0.09 \\ 
Year 19  &  0.157&1.346 $\pm$ 0.484  &42.67  $\pm$   0.15 & 41.13 $\pm$ 0.10 \\ 
Year 20  &  0.227&1.210 $\pm$ 0.409  &42.62  $\pm$   0.14 & 41.15 $\pm$ 0.06 \\ 
\enddata
\tablenotetext{a}{Based on the recalibrated nuclear flux densities at 5100\AA{}, F(5100\AA) \citep{Peterson13}}
\tablenotetext{b}{Uncertainty includes the mean spectral flux measurement uncertainty and the host flux uncertainty. The latter contains an additional 5\% uncertainty due to seeing effects \citep[for details, see][]{Bentz2013}.}
\tablenotetext{c}{Monochromatic nuclear (i.e., host-corrected observed) luminosity at 5100\AA, calculated from Galactic reddening corrected F(5100\AA) values. 
Luminosity errors include the distance uncertainties listed in Table\,\ref{tab:objects}.}
\tablenotetext{d}{$L(\Hbeta)$ luminosities are measured from the mean spectra.}
\label{tab:ngc5548data}
\end{deluxetable*}

We note that in the case of NGC\,5548 for which we have over 20 years of monitoring data,
\citet{Peterson13} do see long term variations in the [\oiii ] line flux and have, based
thereon, re-calibrated the continuum and H$\beta$ flux measurements. We adopt these
new flux values for our study and use the recently updated host galaxy flux measurements 
of \citet{Bentz2013} to compute the corresponding nuclear 5100\AA{} continuum flux 
densities and luminosities for each available monitoring campaign (Table~\ref{tab:ngc5548data}).

\begin{deluxetable}{ccc} 
\tablecolumns{3}
\tabletypesize{\footnotesize}
\tablewidth{0pt}
\tablecaption{Continuum Flux Densities for NGC\,4151 and NGC\,7469}
\tablehead{
\colhead{Object} &
\colhead{JD\tablenotemark{a}} &
\colhead{$F_{cont}$\tablenotemark{b}}}
\startdata
NGC 4151&$49324.0$ &$6.46\pm0.04$\\ 
&$49325.9$ &$6.58\pm0.02$\\
&$49326.9$ &$6.64\pm0.02$\\
&$49327.9$ &$6.69\pm0.02$\\
&$49328.9$ &$6.61\pm0.02$\\
&$49329.9$ &$6.61\pm0.02$\\   
&$49330.9$ &$6.59\pm0.02$\\ 
&$49331.9$ &$6.61\pm 0.02$\\
NGC7469&$50248.5$ &$1.42\pm 0.08$\\
&$50253.5$ &$1.31\pm 0.08$\\
&$50262.7$ &$1.35\pm 0.08$\\
&$50273.5$ &$1.30\pm 0.08$\\
\enddata
\tablenotetext{a}{Julian Dates subtracted by 240000.}
\tablenotetext{b}{ Continuum flux densities at rest frame 5100\AA, in units of $10^{-14}$ erg s$^{-1}$ cm$^{-2}$ \AA$^{-1}$ measured in this work. Host galaxy light is not subtracted.}
\label{tab:Fcont}
\end{deluxetable}

For a couple of the datasets, further processing and/or measurements are required.
For NGC\,4151, the only optical spectra obtained during a UV monitoring campaign are 
those presented by \citet{Kaspi96} from the  $\IUE$ monitoring campaign in 1993.
Among these data, we restrict our consideration to the OSU spectra,   
obtained with the CCDS instrument on the Perkins 1.8-m telescope.
This spectroscopic aperture ($5\arcsec \times 7.5 \arcsec$) is large 
enough to minimize aperture and seeing effects but small enough to enable an accurate star light correction by use of {\it HST} ACS/HRC imaging \citep[e.g.,][]{Bentz2013}.
To perform the absolute flux calibration for this object, we compute the scaling factor to be applied from the observed 
[\oiii]$\lambda 4959$ line emission in order to avoid issues with potential saturation of the [\oiii]$\lambda 5007$ line \citep{Bentz2006}. 
  We adopt the absolute [\oiii]$\lambda 4959$ 
line flux measured by \citet{Bentz2006} from spectrophotometric data.
The continuum flux densities, $F_{cont}$, listed in Table\,\ref{tab:Fcont} are measured 
as the average flux density in the observed reference frame between 5100\AA\ and 5125\AA.
 
Also for NGC\,7469, we use only the OSU subset of the optical dataset presented by 
\citet{Collier98} for data homogeneity reasons.  These data were obtained with the Bollen and Chivens spectrograph 
on the 1.8-m Perkins telescope and a $5\arcsec \times 7.5 \arcsec$ spectroscopic aperture.
For each spectrum we measure the observed continuum flux density, tabulated in 
Table\,\ref{tab:Fcont}, as the mean flux between 5176\AA\ and 5200\AA\ in the observed frame
since \citet{Collier98} measured the continuum at 4845\AA{} that is likely to have \heii\  $\lambda$4686 and \feii{} contamination.

\subsubsection{Reddening Correction}\label{S:2.1.2}

We correct the optical and UV continuum flux densities for extinction due to the  
Galaxy using the extinction curve of  \citet{Cardelli89} and the $E(B-V)$ values relevant 
for each source based on the \citet{Schlafly2011} recalibration of the dust maps of  
\citet{Schlegel98} as listed in the NASA/IPAC Extragalactic Database (Table\,\ref{tab:objects}). 
With no robust way to estimate the nature and amount of the dust extinction of the intergalactic medium 
between the AGN and us or the interstellar medium of the AGN host galaxy,  
we do not apply any correction for these two potential sources of extinction. 
However, internal reddening is typically expected to be rather low in these objects,
as we do not observe a UV-optical spectrum deviating strongly from a power-law
\citep[e.g.,][]{Crenshaw2001, Richards2003}. 
Therefore, we do not expect the lack of internal dust correction to adversely affect our 
analyses and results. 

\subsubsection{Host Galaxy Star Light Correction}\label{S:2.1.3}

We subtract the star light flux measured by \citet{Bentz2013} for each AGN from
the observed optical flux density to obtain the {\it nuclear} luminosity $L(5100\AA)$, adopting   
the host galaxy flux measured for the same specific aperture size and position
as was used for the spectroscopic observations; 
these are listed in Table~\ref{tab:objects}.
We note that NGC\,7469 has a nuclear star-forming ring with a diameter of $\sim$5\arcsec{} 
that is visible in both optical and UV imaging \citep{DiazSantos2007}. 
The optical fluxes from this spatially resolved star-forming ring 
is included in the host flux measurements adopted here, 
but the UV luminosities are not corrected for the contribution from young stars 
in the starburst ring. 

\subsection{Data for the $R-L(5100\AA)$ Relationship for Seyfert 1 Galaxy NGC\,5548}\label{S:2.2}

For our analysis of the native $R(\Hbeta) - L(5100\AA)$ relationship for NGC\,5548 we
use the results of 15 individual monitoring campaigns that each provide independent
measurements of $R(H\beta)$ for a given $L(5100\AA)$.
NGC\,5548 was monitored for 13 years by the International 
AGN Watch program \citep{Peterson2002} starting in 1988, and again in 2005 \citep[`Year 17';][]{Bentz2007}, 
2007 \citep[`Year 19';][]{Denney10}, and 2008 \citep[`Year 20';][]{Bentz09b}.
Collectively, these campaigns provide 16 individual measurements of $R(\Hbeta)$ for various 
luminosity states spanning more than 20 years. 
However, we exclude `Year\,19' 
because the results were somewhat 
ambiguous: the cross-correlation function (CCF)  is broad and flat-topped with a ``maximum'' ranging from $\sim$3 to $\sim$23 days \citep[Figure~3 of][]{Denney10}. The velocity resolved 
time delay \citep[Figure~4 of][]{Denney10} corroborates  that the \Hbeta{} 
emitting line region responds at this range of time scales. Although this may be real, such 
a broad range of possible lags for a single epoch does not provide sufficient information 
to be useful here. 
For the other 15 campaigns, we compute the nuclear AGN luminosities (listed in Table\,\ref{tab:ngc5548data}) based on the most
recent recalibration of the [\oiii]\,$\lambda$5007 narrow-line flux of NGC\,5548 \citep{Peterson13}, the most 
recently updated host-galaxy flux measurements \citep[][]{Bentz2013},  
and the source  distance listed in Table~\ref{tab:objects}.

NGC\,5548 is the only object for which we can generate a native $R(\Hbeta) - L(5100\AA)$  
relationship because other objects in the RM sample have only been monitored during a single reverberation mapping campaign, or at most a couple of campaigns, insufficient for this study.  Because of this, as well as the fact that intrinsic variability drives this single-object $R(\Hbeta) - L(5100\AA)$ relationship, we first verify that NGC\,5548 is representative of other objects in the RM sample, with respect to variability properties.
In the histogram in Figure~\ref{fig:hts} we compare the fractional variation, 
$F_{var}$(continuum), a measure of the intrinsic variation amplitudes of the nuclear continuum
\citep{Rodriguez-Pascual97}, for all the sources in the RM sample based on the previous
published RM studies. 
We computed the $F_{var}$ values for NGC\,5548 (listed in Table~\ref{tab:ngc5548data}, column 2) based on equation~(3)
of \citet{Rodriguez-Pascual97} using the updated host fluxes and Galactic reddening of 
\citet{Bentz2013} and the re-calibrated measured flux densities in column 3 of  Table\,\ref{tab:ngc5548data}. 
Since the $F_{var}$(continuum) values for NGC\,5548 (gray shaded histogram) 
fall in the middle of the sample distribution, it is reasonable to assume for the 
purpose of this investigation that NGC\,5548 is representative of the RM sample. 
Yet, this comparison also shows that NGC\,5548 does not probe the most extreme 
variability of the RM sample, which is about 50\% larger than that of NGC\,5548. 
\citet{sergeev07} present light-curves from 30 years of monitoring 
NGC\,5548 from 1972 $-$ 2001 and find similar variability characteristics during this 
period and when comparing the earlier (1972 $-$ 1988) and later (1989 $-$ 2001) campaigns. 
This demonstrates that the 20 year period over which our observations span is 
representative of all known variability characteristics of this source.

\setcounter{figure}{0}
\begin{figure} 
\begin{center}$
\begin{array}{c}
\includegraphics[scale=0.35,angle=270]{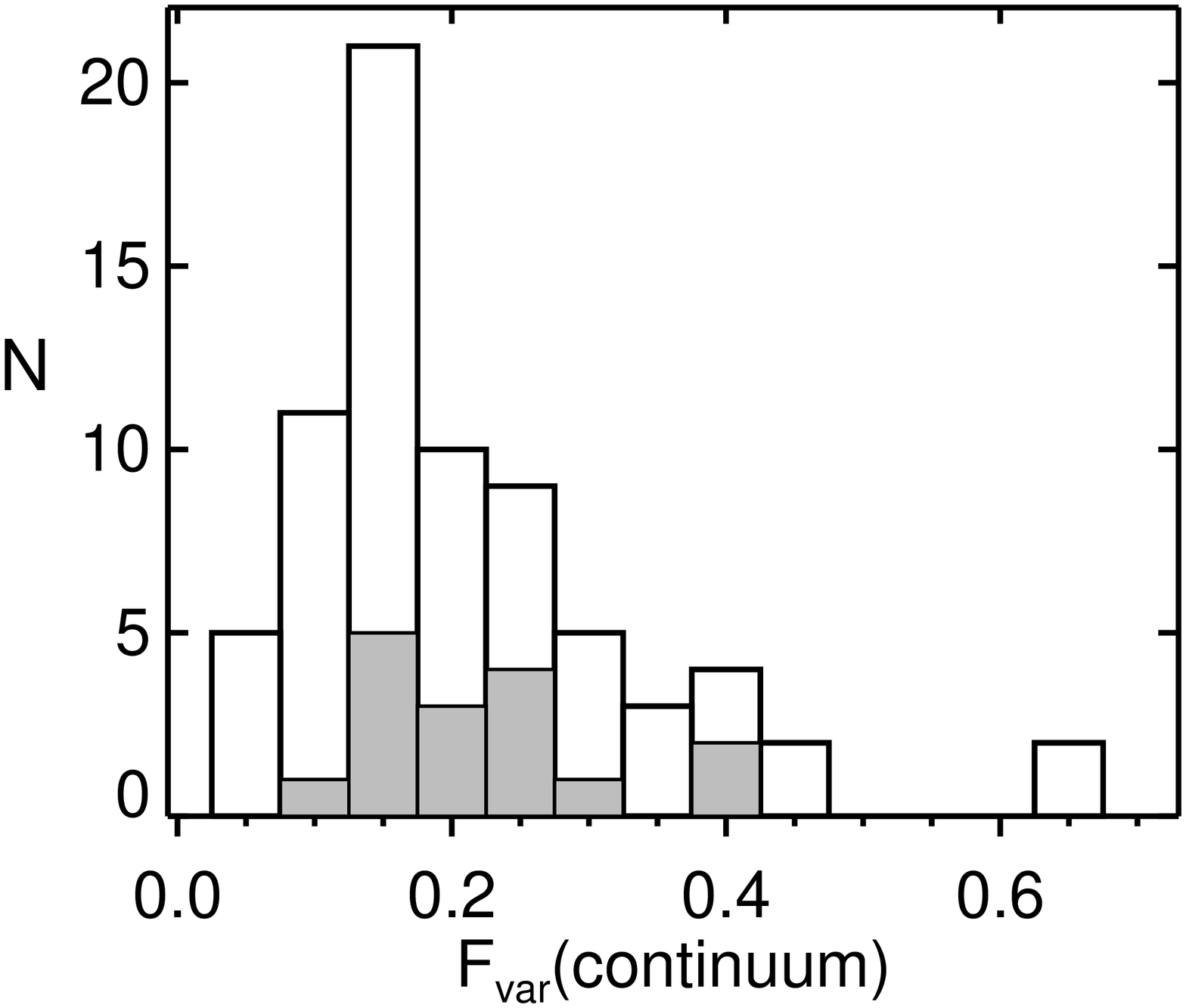}\\  
\end{array}$
\end{center}
\vspace{0.2cm}
\caption{Histogram of the $F_{var}$(continuum) distribution for the full RM sample of \citet{Bentz2013}. 
The $F_{var}$(continuum) distribution for NGC\,5548, shown with a gray shade, is seen to be representative of
the RM sample.
The $F_{var}$ values are adopted from \citet{Peterson04, Denney2006, Bentz09b, Denney10, Grier2012, Peterson13} 
and corrected for the host galaxy contribution to the spectrally measured 
monochromatic source luminosities.} 
\label{fig:hts}
\end{figure}  

We use the results of two different analysis methods to determine the $R(\Hbeta)$ values because they yield different uncertainties that can affect the scatter that we aim to quantify. 
The cross-correlation function (CCF) method uses cross-correlation of the intra-day interpolated continuum and emission-line light curves to determine the time delay \citep[see][for details]{Peterson04}, while `JAVELIN\footnote{JAVELIN (`Just Another Vehicle for Estimating Lags In Nuclei') is formerly known as `SPEAR'; https://bitbucket.org/nye17/javelin.}' \citep{Zu2011} uses more advanced statistical Markov Chain Monte Carlo techniques to derive the delay, taking advantage of the observation that AGN variability can be well described as a damped random walk process \citep{Kelly2009,Kozlowski2010,MacLeod2010}.
The 15 $R(\Hbeta)$ measurements based 
on the CCF analysis are adopted from \citet{Peterson04}, \citet{Bentz2007}, and \citet{Bentz09b}. 
The JAVELIN analysis of \citet{Zu2011} provides 13 of the 15 measurements of $R(\Hbeta)$, since 
that study did not include the \citet{Bentz2007} and \citet{Bentz09b} data.
To allow a direct comparison with the CCF database we add our own, similar, JAVELIN analysis \citep[following][]{Grier2012} of these two campaigns for which we obtain $\tau_{\rm rest}$ = $5.54^{+2.32}_{-1.85}$ days for year 17 \citep{Bentz2007} and $\tau_{\rm rest}$ = $4.52^{+0.36}_{-0.33}$\,days for year 20 \citep{Bentz09b}.

\section{Analyses and Results}\label{S:S3}

\begin{figure} 
\begin{center}$
\begin{array}{c}
\includegraphics[scale=0.6,angle=0]{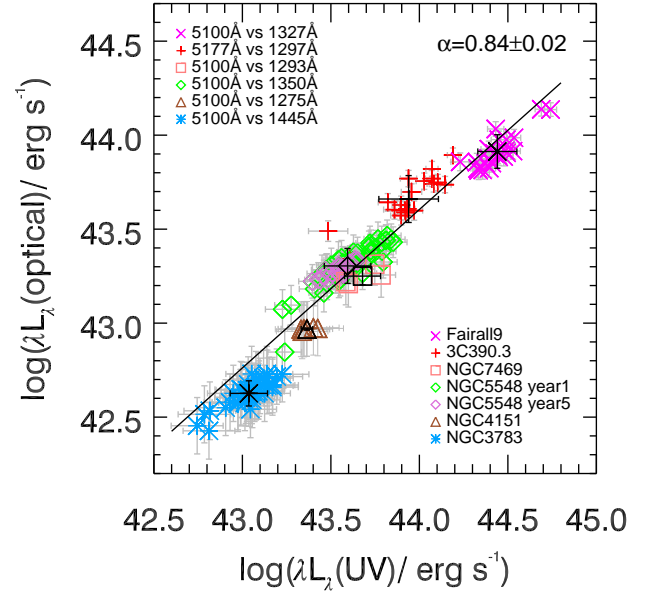}\\
\end{array}$
\end{center}
\vskip -0.3cm
\caption{Optical continuum luminosity versus   
ultraviolet continuum luminosity for the six nearby AGN in our sample; the optical 
and UV measurements are paired to within two days. The optical luminosities 
are corrected for the host galaxy star light entering the spectroscopic aperture. 
The luminosities are 
measured at the specific wavelengths for each source as listed in the upper left 
corner of the diagram. Black points denote the mean luminosities of the AGN 
and their error-bars represent the 1\,$\sigma$ standard deviation of the 
luminosities for each object.  The solid line shows the (global) best-fit to all sources.
While a global linear relation is seen for the full sample, individual AGN exhibit native
relationships that are slightly offset from the global relationship and have shallower slopes. 
}
\label{fig:Lrel}
\end{figure}

\begin{deluxetable*}{lrccc} 
\tablecolumns{5}
\tabletypesize{\footnotesize}
\tablewidth{0pt}
\tablecaption{Regression Results for the Optical-UV Luminosity Relationships} 
\tablehead{
\colhead{Object} & 
\colhead{Zeropoint, $A$\tablenotemark{a}} & 
\colhead{Slope, $\alpha$\tablenotemark{a}} &
\colhead{$\sigma_{RMS}$\tablenotemark{b}} &
\colhead{$\epsilon_{0}$\tablenotemark{c}}\\
\colhead{} &
\colhead{} &
\colhead{}&
\colhead{(dex)}&
\colhead{(dex)}\\
\colhead{(1)} &
\colhead{(2)} &
\colhead{(3)} &
\colhead{(4)} &
\colhead{(5)}}
\startdata
Fairall\,9                                &                $ 0.56\pm0.06$     &$0.78\pm0.13$             & 0.049 &0.027$\pm$0.014  \\
3C\,390.3                              &                $ 0.70\pm0.02$     &$0.84\pm0.14$             & 0.074 &0.037$\pm$0.018 \\
NGC\,7469                            &                 $0.36\pm0.25$   &0.40$\pm0.23$            & 0.038 & 0.035$\pm$0.034 \\
NGC\,5548 All                       &                 $ 0.55\pm0.05$    &$0.63\pm0.12$            & 0.043 & 0.017$\pm$0.008 \\
NGC\,5548 year1                   &                 $ 0.56\pm0.05$     &$0.65\pm0.14$           & 0.050 & 0.019$\pm$0.012 \\
NGC\,5548 year5                   &                 $ 0.45\pm0.10$     &$0.39\pm0.22$                                & 0.020 & 0.025$\pm$0.015 \\  
NGC\,3783                            &                 $ 0.15\pm0.05$    &$0.54\pm0.05$                                 & 0.036 & 0.037$\pm$0.004 \\ 
NGC\,4151                            &                 $ 0.04\pm0.05$    &$0.12\pm0.09$                                  & 0.005& 0.006$\pm$0.004 \\
All Sources                          &                   $ 0.61\pm 0.01$      &$0.84\tablenotemark{d}\pm0.02$& 0.114 &0.055$\pm$0.010 \\
All Sources                          &                   $ 0.63\pm 0.01$      &[1.00\tablenotemark{e}$\pm$0.00]& 0.109 &0.089$\pm$0.009 \\ 
All Sources except NGC\,7469 and NGC\,4151& $0.61\pm 0.01$      &$0.84\tablenotemark{d}\pm0.02$& 0.117 &0.054$\pm$0.001 \\ 
All Sources except NGC\,7469 and NGC\,4151& $0.63\pm 0.01$      &[1.00\tablenotemark{e}$\pm$0.00]& 0.111 &0.091$\pm$0.010 \\
{Mean Luminosities\tablenotemark{f}} &     $ 0.59\pm 0.11$      &0.96$\pm$0.21& 0.804 &0.121$\pm$0.129 \\
\enddata
\tablenotetext{a}{Best fit parameters for 
$\log [\lambda L_{\lambda}{\rm (optical)}] =A +\alpha \times \log [\lambda L_{\lambda}{\rm (UV)}]+\epsilon_{0}$. 
The parameters are the median values of the posterior probability distributions, while the uncertainties are the standard deviation of the posterior distributions. } 
\tablenotetext{b}{The RMS scatter relative to the listed relationship. }
\tablenotetext{c}{The measured scatter: the square-root of the median of the posterior probability distribution of 
the variance of the  scatter. } 
\tablenotetext{d}{This slope of the measurements for the entire source sample is referred to in 
the text as `the global slope'.}
\tablenotetext{e}{The slope is held fixed to unity during the regression to allow a measure of the  scatter.}
\tablenotetext{f}{The mean luminosities, shown as black symbols in Figure \ref{fig:Lrel}. 
}
\label{tab:lopluvlinmix}
\end{deluxetable*}

\begin{deluxetable*}{lcccc} 
\tablecolumns{5}
\tabletypesize{\footnotesize}
\tablewidth{0pt}
\tablecaption{Light Curve Statistics of the Optical-UV Database}
\tablehead{
\colhead{Object} &
\colhead{Simultaneity Time Frame\tablenotemark{a}} &
\colhead{$N$\tablenotemark{b}} & 
\colhead{$F_{var}$(optical)} & 
\colhead{$F_{var}$(UV)} \\
\colhead{}&
\colhead{(days)} &
\colhead{} &
\colhead{} &
\colhead{}\\
\colhead{(1)} &
\colhead{(2)} &
\colhead{(3)} &
\colhead{(4)} &
\colhead{(5)}}
\startdata
NGC\,4151                             &17     &8     &0.012 & 0.077\\
NGC\,5548 year5                    &37     &24   &0.056&0.158\\
NGC\,7469                             &25     & 4    &0.080&0.229\\
NGC\,3783                             &221   &44   &0.142& 0.218\\
NGC\,5548 All                        &1624  &75   &0.176&0.291\\
NGC\,5548 year1                     &236   &51   &0.197&0.302\\
Fairall\,9                               &188   & 24    &0.233&0.287\\
3C\,390.3                              &333   &20    &0.268&0.319\\ 
\enddata
\tablenotetext{a}{The time span covered by the simultaneous optical and UV data analyzed here.}
\tablenotetext{b}{Number of optical-UV data pairs.}
\tablecomments{The entries in this Table pertain only to the subset of RM AGN analyzed in Figure\,\ref{fig:Lrel} for the listed subset of monitoring data. The $F_{var}$ histogram in Figure\,\ref{fig:hts} is based on the full dataset of monitoring data for the full sample of RM AGN.}
\label{tab:fvar}
\end{deluxetable*}

\subsection{The Optical$-$UV Continuum Luminosity Relationship } \label{S:3.1}

We investigate the relationship between multiple epochs of simultaneous measurements 
of optical and UV continuum luminosities for individual sources and for the 
sample as a whole.  The goals are to establish whether the optical and UV luminosities
are mutually interchangeable and, if not, to estimate how much scatter can be introduced
into the radius $-$ luminosity relationship by adopting $L$(5100\AA) rather than  $L$(UV)
as a proxy for $L$(ionizing).
For this analysis, we compile quasi-simultaneous measurements (within two days) 
of $L$(optical) and $L$(UV), as described in \S\ref{S:2.1}.  
We show the relationship between the quasi-simultaneous optical and UV luminosities for the RM sub-sample in Figure\,\ref{fig:Lrel}, where each object is identified by its own symbol. For NGC\,5548 we have two datasets obtained during two different 
monitoring campaigns. 
`NGC\,5548 year 1' (`NGC\,5548 year 5') refers to the monitoring campaign that 
ran in 1988 (1993).     
We refer to the combined dataset of years 1 and 5 as 'NGC\,5548 All'.
A clear, positive trend between $L$(optical) and $L$(UV) is seen in Figure\,\ref{fig:Lrel}, as
expected. Yet, we find that each individual object exhibits its own, i..e., `native',
$L$(optical) $- L$(UV) correlation that differs in slope from that of other objects and 
also from the `global' relationship that exists across the entire sample.

To characterize the $L$(optical)$-$$L$(UV) relationship, we adopt the following parameterization:
\begin{equation}\label{E:linmixerr} 
\log \Bigg[\frac{\lambda L_{\lambda}{\rm (optical)}}{10^{43}\,{\rm erg\, s}^{-1}}\Bigg] = A +\alpha \log\Bigg[\frac{\lambda L_{\lambda}{\rm (UV)}}{10^{44}\,{\rm erg\, s}^{-1}}\Bigg] +\epsilon_{i}.
\end{equation}

\noindent where $A$ is the zero point, $\alpha$ is the slope, and $\epsilon_{i}$ is the estimated 
scatter\footnote{Note that while mathematically this is often referred to as the `intrinsic scatter' 
(i.e., the additional scatter required, above that accounted for by the measurement errors, so that the regression analysis produces a $\chi^2$ value of 1.0), by the nature of the observations we do not know if this scatter really is intrinsic or contains contributions from uncertainties in measurements and flux corrections. Therefore, we will refer to this scatter as the `estimated scatter'.}. We establish the best fit relationship for each object 
and the sample as a whole by use of the Bayesian regression method\footnote{implemented in IDL as `LINMIX$\_$ERR.pro'} of  
\citet{Kelly2007} because it is more robust than the commonly used FITEXY
$\chi^{2}$ minimization method  \citep{Press92} for small samples.  
The Bayesian method accounts for measurement uncertainties in both variables and the 
scatter, $\epsilon_{i}$, and computes the posterior probability distributions of the parameters in 
Equation~(\ref{E:linmixerr}). 
This method uses Gaussian distributions to describe the measurement errors and the 
scatter, and a `Gaussian mixture model' to represent the distribution of the independent variable. 
Since our dataset is relatively small, we use only a single Gaussian in the `mixture 
modeling' to speed up the computations.

The results of our regression analysis are listed in Table\,\ref{tab:lopluvlinmix}, which 
contains the source name (column 1), 
the best fit parameters: intercept and slope (columns 2 and 3, respectively), the rms scatter,
  $\sigma_{RMS}$, of the data relative to the individual best fit relationships 
(column 4), and the estimated scatter, $\epsilon_{i}$ (column 5). 
Light curve statistics for the six AGN in our sample are listed in Table~\ref{tab:fvar} with the time span for which we have simultaneous optical and UV data in column 2, the number of data pairs (epochs) in column 3, and the $F_{\rm var}$ values (\S\,\ref{S:2.2}) for the optical and UV continuum light curves in columns 4 and 5, respectively.

In Figure \ref{fig:Lrel}, we also show the best fit relationship for the entire 
sample (i.e., the global fit; solid line) by taking into account all the individual data points.
The global $L$(optical) $- L$(UV) relationship has a slope $\alpha=0.84\pm0.02$, while 
the slope is different for each individual AGN, with values in the range 
between 0.12 to 0.84 (Table\,\ref{tab:lopluvlinmix}).  With these single-object slopes being different from
unity, there is not a one-to-one correspondence between the two luminosities for most of the AGN;
only for Fairall\,9 and 3C\,390.3 are the measured slopes consistent with unity to within 2\,$\sigma$.
The $F_{var}$(UV) values are clearly all larger than $F_{var}$(optical), showing stronger variability amplitudes at UV energies, as also indicated by the shallow $L$(optical)$-L$(UV) slopes.

NGC\,7469 and NGC\,4151 exhibit somewhat shallower slopes and lower variability amplitudes than the other AGN in the sub-sample. This is likely related to the very few available data points, obtained over a brief time span. Although these data are not likely to be representative of the intrinsic variability properties of these two AGN over similar time scales as that covered by the observations of the rest of this sub-sample, omitting these datasets do not change the results (Table\,\ref{tab:lopluvlinmix}).

Figure\,\ref{fig:Lrel} shows that the single-object $L$(optical)$-L$(UV) relationships do not fall on top of the best fit to the global relationship (solid line in Figure\,\ref{fig:Lrel}) but instead show slight zero-point offsets. These offsets can be due to,
e.g., (1) intrinsic differences in the spectral energy distributions between objects, (2) imperfect host galaxy flux subtraction, (3) imperfect absolute spectrophotometric calibration, which was performed differently for the optical and UV data, or (4) uncorrected internal dust reddening in the AGN host galaxy that will be different for each object.
Note that while the former two effects can impact both the zero-point and slope of the native  $L$(optical) $-$ $L$(UV) relationship, the latter two will not affect the slope of the native relationship. 
Combined with the single-object (i.e., native) slopes being shallower than the global slope, this introduces a scatter in the global relationship. 
Our Bayesian analysis estimates the scatter of all the data pairs relative to a unity global relationship to be $\epsilon_0$ = 0.09\,dex. 
Given the relatively short time scales for which we have quasi-simultaneous optical and UV luminosities, this scatter must represent a lower limit of the scatter we can expect by our use of $L$(optical) rather than $L$(ionizing) for the luminosity in the global $R(\Hbeta) - L$(ionizing) relationship.

\begin{figure}[th] 
\begin{center}$
\begin{array}{c}
\includegraphics[scale=0.45]{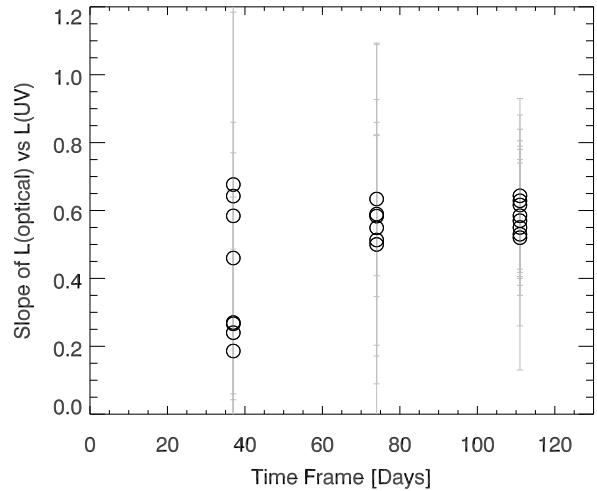}\\ 
\end{array}$
\end{center}
\caption{Dependence of the NGC\,5548 $L$(Optical) $-$ $L$(UV) relationship slope on time span of the data. From the Year\,1 dataset with a time span of 236 days, datasets with shorter time spans of 37 days, 74 days, and 111 days were selected and the slopes were measured. There is a tendency for a large range of measured slope for the datasets with short time spans compared to datasets with longer time spans.}
\label{fig:figB1}
\end{figure}

One effect that can explain object-to-object differences is the accretion state of the central engine. 
Depending thereon, the specific $L$(optical)$-L$(UV) relationship may change significantly in 
time: in the case of NGC\,5548, the slope changes from 0.65 (year 1) to 0.39 (year 5). 
However, since the `year\,5' dataset only covers 37 days, while `year\,1' covers 236 days, the shallower slope may also be related to the time span over which we have simultaneous optical and UV data, in this case. This is confirmed in Figure~\ref{fig:figB1}: 
we selected from the `year\,1' dataset  subsamples in multiples of 37 days (i.e., 37 days, 74 days, and 111 days, respectively; the time span is not necessarily of contiguous days as time gaps exist) and measured the slope for each data subset. We see a clear tendency for a wider range of slopes (between 0.18 and 0.66) for subsets of 37 days than for subsets of 111 days (slope\,$\sim$0.6).
These results demonstrate that since AGN continuum variations are unpredictable, when the slope is determined from data covering relatively short time spans, it can generally not be assumed to be valid at other times. In that case,
the slope distribution is more likely to exhibit a larger dispersion because the data capture individual, shorter time-scale variability events (that can include either large amplitude changes or none at all), typical of Seyferts observed over days to weeks.  On the other hand, the slope measured from data covering a longer time span is more likely to represent the variability characteristics observed over timescales of months to years (often referred to as `secular'). The long-term variations likely correlate with the mean accretion state, which is typically much more stable, over dynamical timescales, or longer.

Another potential effect that can alter the slopes of the native, single-object  $L$(optical)$-L$(UV) relationships is if the host galaxy flux contribution, $F_{host}$ (Table\,\ref{tab:objects}, column 5), is mis-estimated.
While the risk is low since we adopt the well-determined host flux measurements of \citet{Bentz2013} based on high spatial resolution and high signal-to-noise {\it HST} and ground-based imaging, we test this possibility nonetheless.
Given the shallow slopes, the only way to get a linear relationship is to assume that  the host galaxy flux level of each object is underestimated.  
To test this, we iteratively subtract an increasing amount of host galaxy flux in addition to that listed in Table~\ref{tab:objects} until we measure a slope of 1.0 for each source.  
We list this additional amount of host galaxy flux needed as $F_{gal,extra}$ in Table~\ref{tab6}. 
We find $F_{gal,extra}/F_{host}$ ratios in the range between $\sim$20\% and $\sim$300\%, and for most of the sources $F_{gal,extra}$ is a significant fraction of $F_{host}$ (35\% or more; Table~\ref{tab6}). 
These $F_{gal,extra}$ values correspond to 4$ \sigma_{host} - $ 30 $\sigma_{host}$, a statistically significant change. 
Furthermore, the value of $F_{gal,extra}$ estimated for NGC\,5548 is particularly unrealistic because if we were to subtract this extra flux from the continuum measurements presented by \citet{Peterson13}, the continuum fluxes would be negative for some epochs. 
We consider it unlikely that the host galaxy flux would be so grossly underestimated and
conclude that the observed optical and UV variability amplitude differences cannot be attributed to an inaccurate correction for host galaxy flux contamination at optical wavelengths.

\begin{deluxetable}{lccc} 
\tablecolumns{4}
\tabletypesize{\footnotesize}
\tablewidth{230pt}
\tablecaption{Host Galaxy Flux Density Comparison}
\tablehead{
\colhead{Object} &
\colhead{$F_{gal,extra}$\tablenotemark{a}} &
\colhead{$F_{gal,extra}$} &
\colhead{$F_{gal,extra}$}\\
\colhead{} &
\colhead{} &
\colhead{/$\sigma_{host}$\tablenotemark{b}} &
\colhead{/$F_{host}$\tablenotemark{b}}\\
\colhead{} &
\colhead{(erg s$^{-1}$ cm$^{-2}$ \AA$^{-1}$)} &
\colhead{}&
\colhead{(\%)}\\
\colhead{(1)} &
\colhead{(2)} &
\colhead{(3)} &
\colhead{(4)}}
\startdata
Fairall 9           &$6.60\times 10^{-16}$&~~4.4   &~~22\\  
3C 390.3          &$1.73\times 10^{-16}$&~~4.2   &~~21\\ 
NGC 7469         &$3.00\times 10^{-15}$&~~3.8   &~~35\\  
NGC 5548 All    &$2.48\times 10^{-15}$&~~6.6   &~~65\\ 
NGC 3783        &$2.64\times 10^{-15}$&~5.6    &~~55\\ 
NGC 4151        &$4.24\times 10^{-14}$&30.3    &282\\ 

\enddata
\tablenotetext{a}{ The amount of additional host galaxy flux density needed to obtain a slope of one in the $\lambda L_{\lambda}({\rm optical})$-$\lambda L_{\lambda}({\rm UV})$ relationship for each AGN.} 
\tablenotetext{b}{
The values of the host galaxy flux, $F_{host}$, and the measurement uncertainty, $\sigma_{host}$, 
are listed in Table\,\ref{tab:objects} and are adopted from \citet{Bentz09a} and \citet{Bentz2013}. }
\label{tab6}
\end{deluxetable}

\subsection{The Radius$-$Luminosity Relationships of NGC\,5548} \label{S:3.3}

Our second study addresses the contribution to the observed scatter in the global 
$R(\Hbeta) -L$(5100\AA) relationship (i.e., based on the full sample of RM AGN) 
from the scatter introduced by a {\it single} object as it varies in luminosity over time. 
Longer term variations over time scales of several years will better probe this scatter
as each measurement on the global relationship was obtained at a random time during 
the lifetime of each AGN. 
Ideally, we would want to examine how $R(\Hbeta)$ changes with the UV luminosity for 
all the objects in the RM sample, as $L(UV)$ is expected to be a better estimate 
than $L(5100\AA)$ of the ionizing luminosity that dictates the size of the $\Hbeta$ 
emitting region. Unfortunately, only for one object, NGC\,5548, can the available data address its long term variability properties.
Because the data are not available for a detailed analysis of the $R(\Hbeta)-L$(UV)
relationship itself for NGC\,5548, we examine instead the $R(\Hbeta)-L$(5100\AA) 
relation and the implications for its scatter from the observed 
$L$(5100\AA) $- L$(UV) relationship, presented above. We then examine the inferred
$R(\Hbeta)-L$(1350\AA) relationship in order to test our assumption that $L$(UV) is 
a better proxy for $L$(ionizing).

\subsubsection{The $R-L(5100\AA)$ Relationship for NGC\,5548 }\label{S:3.3.1} 

\begin{figure} 
\begin{center}$
\begin{array}{c}
\includegraphics[scale=1.0]{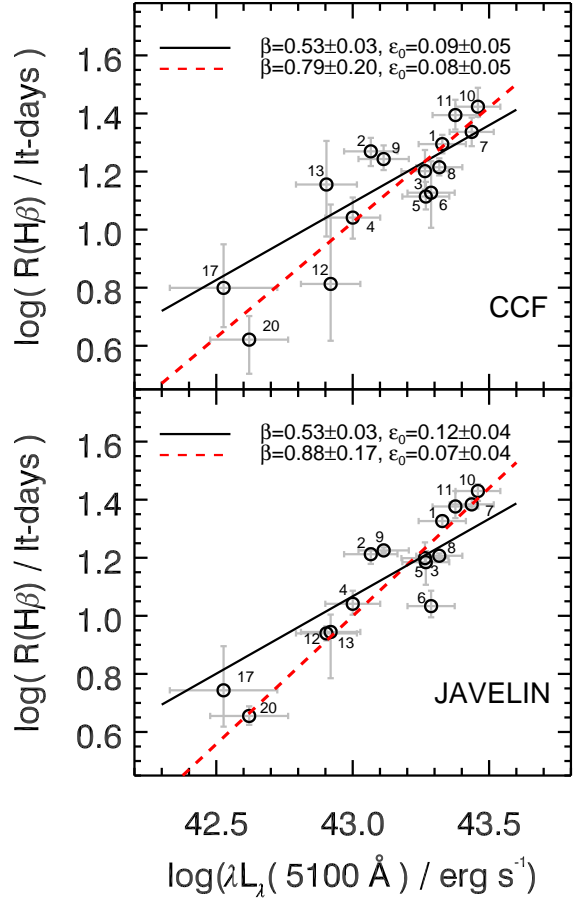}\\
\end{array}$
\end{center}
\vspace{-0.5cm}
\caption{The $R-L(5100\AA)$ relationships for NGC\,5548 based on the CCF dataset (top) 
and the JAVELIN dataset (bottom) updated to account for the new flux calibration of \citet{Peterson13}. The numbers refer to the year of the reverberation 
mapping campaign as described in \S\ \ref{S:2.2}. 
The red dashed lines in each panel show the best fit relationship to each dataset (the relation traced by the intrinsic variability of NGC\,5548).  
The black solid lines show the global relationship with slope $\beta=0.53$ \citep{Bentz2013}.
}
\label{fig:rl1}
\end{figure}

Figure~\ref{fig:rl1} shows the $R(\Hbeta)-L$(5100\AA) relationships for the CCF (top) 
and JAVELIN (bottom) datasets.  The red dashed lines are the best fit regressions to each dataset. 
We describe the $R-L(5100\AA)$ relationship as:
\begin{equation}\label{E:rlngc5548} 
\log \Bigg[\frac{R_{BLR}}{\rm 1\,light-day}\Bigg] = K +\beta \log\Bigg[\frac{\lambda L_{\lambda}(5100\AA)}{{10^{44}\, \rm erg\, s}^{-1}}\Bigg] +\epsilon_{0},
\end{equation} 
\noindent 
where $K$ is the zero point, $\beta$ is the slope, and $\epsilon_{0}$ is the estimated scatter. 
Because the regression method cannot account for the asymmetric 
uncertainties in our $R_{BLR}$ measurements we performed an extensive
`error-bar sensitivity test' (described in Appendix~\ref{S:AppendixA}) to test 
the effects of adopting a particular symmetric uncertainty on $R_{BLR}$.
The test revealed that the regression results are not significantly 
affected by which of the possible error-bars we adopt.  
To be conservative we adopt the larger of the upper and lower $1\sigma$ 
uncertainties for each object and quote the best fit parameters to equation (\ref{E:rlngc5548}) based thereon. 
The best fit slope and intercept obtained from 
the Bayesian analysis are the median values of the posterior probability 
distributions while the quoted uncertainties are the standard deviations 
with respect to the median.
For each of the CCF and JAVELIN datasets, Table~\ref{tab7} lists the resultant zero-point and slope of the
$R-L(5100\AA)$ relationships (columns 2 and 3, respectively);
the root mean square scatter, $\sigma_{RMS}$, 
of the $R_{BLR}$ data (column\,4) relative to the best fit relationship; 
the estimated scatter $\epsilon_{0}$ (column\,5); and the precision of the scatter estimate
(column\,6).  

The regression slopes obtained from both datasets agree to within the errors.  
This is expected since \citet{Zu2011} found mostly consistent lag measurements 
($R_{BLR}$) for the CCF and JAVELIN analysis methods. 
However, the CCF dataset has larger $R_{BLR}$ uncertainties.
\citet{Bentz2007} examine the $R(\Hbeta)-L$(5100\AA) relationship for NGC\,5548 using 
only the CCF data of the first 14 campaigns (Year 1\,$-$\,17) and find a slope 
$\beta=0.73\pm0.14$. 
\citet{Zu2011} examine the same relationship with the 
JAVELIN dataset and obtain a slope $\beta=0.73\pm0.10$.  
We note that our current dataset is somewhat improved compared to these studies owing to updates to the host galaxy contribution measured for individual spectral apertures and Milky Way reddening corrections \citep{Bentz2013} and the improved calibration of the nuclear fluxes \citep{Peterson13}. As a result, we obtain slightly steeper slopes ($\beta=0.79\pm0.20$ for the CCF data; $\beta=0.88\pm0.17$ for the JAVELIN data) than these previous studies, but our results are still consistent to within the uncertainties.

\begin{deluxetable*}{cccccc} 

\tablecolumns{6}
\tabletypesize{\footnotesize}
\tablewidth{0pt}
\tablecaption{ Regression Results for NGC 5548} 
\tablehead{
\colhead{Type of } &
\colhead{Intercept $K$\tablenotemark{a}} &
\colhead{Slope $\beta$\tablenotemark{a}} &
\colhead{$\sigma_{RMS}$\tablenotemark{b}}&
\colhead{$\epsilon_{0}$\tablenotemark{a}}&
\colhead{$\Delta\epsilon_{0}$\tablenotemark{c}}\\
\colhead{Relationship} &
\colhead{(dex)} &
\colhead{}&
\colhead{(dex)} &
\colhead{(dex)} &
\colhead{(dex)}\\
\colhead{(1)} &
\colhead{(2)} &
\colhead{(3)} &
\colhead{(4)} &
\colhead{(5)} &
\colhead{(6)}}
\startdata
\multicolumn{6}{c}{\boldmath{$R -L(5100\AA)$} \bf Relationship}\\
\cutinhead{CCF $R_{BLR}$} 
Native relationship    &$1.83\pm0.15$   &0.79$\pm$0.20   & 0.113        &  0.076             & $\pm$0.047\\ 
Global relationship  &$1.62\pm0.04$  &[0.53$\pm$0.03]   & 0.122      &  0.087       & $\pm$0.046\\   
\cutinhead{JAVELIN $R_{BLR}$} 
Native relationship     &$1.89\pm0.13$    &$0.88\pm0.17$   &0.092       &0.071    &$\pm0.039$ \\     
Global relationship   &$1.59\pm0.04$    &[0.53$\pm$0.03] &0.103        &0.118    &$\pm0.041$ \\[5pt] 
\hline \hline \\[-3pt]
\multicolumn{6}{c}{\boldmath{$R -L(1350\AA)$} \bf Relationship}\\ 
\cutinhead{CCF $R_{BLR}$} 
Native relationship      &$1.48\pm0.10$ &0.47$\pm$0.15   & 0.115    &0.081             &$\pm$0.052\\
Global relationship    &$1.52\pm0.04$ &[0.53$\pm$0.03]  & 0.126   &0.072             &$\pm$0.050\\
\cutinhead{JAVELIN $R_{BLR}$} 
Native relationship      &$1.45\pm0.05$ &0.52$\pm$0.12    &0.105     & 0.073             &$\pm$ 0.052\\
Global relationship    &$1.51\pm0.03$ &[0.53$\pm$0.03]   &0.096   & 0.065             &$\pm$ 0.041
\enddata
\tablecomments{
The values in square brackets \citep[slope $\beta$; ][]{Bentz2013} are held fixed during the 
regression in order to estimate the scatter relative to this particular slope. The zero-point is the best fit value given the data and the adopted slope.
The UV luminosities are computed as described in \S~\ref{S:3.3.2}.}
\tablenotetext{a}{Best fit parameters for the relationship in Equation (\ref{E:rlngc5548}) 
based on adopting the larger of the two error-bars; 
this is option (d) of the `error bar sensitivity test', 
described in Appendix~\ref{S:AppendixA}. 
These parameters and their uncertainties are the median and standard 
deviation of the posterior probability distributions. }
\tablenotetext{b}{The rms scatter of the data points relative to the best fit relationships.}
\tablenotetext{c}{The standard deviation of the posterior probability distribution of the  scatter, i.e., the precision of the scatter estimates.}
\label{tab7} 
\end{deluxetable*}

The best fit slope for each of the CCF and JAVELIN dataset is steeper than the global slope, $\beta$ = 0.53, established by \citet{Bentz2013}.
For the CCF dataset, the uncertainty on the slope shows that there is less than a 20\% probability that the native slope is intrinsically similar to the global one and, therefore, we consider this steeper native slope to be real.
Since we argue that NGC\,5548 is representative of reverberation mapped AGN (\S\,\ref{S:2.2}), a steeper native slope is likely typical of AGN. This suggests that the intrinsic variability of individual sources introduces additional scatter into the global relationship.  
We are, therefore, interested in assessing the scatter on the global relationship introduced by this 
particular well-studied object. 
We estimate the scatter of the NGC\,5548 $R(\Hbeta)$ measurements relative to 
the global relationship (black solid lines in Figure~\ref{fig:rl1})
by fitting each of the CCF and JAVELIN datasets with a fixed slope of $\beta=0.53\,\pm\,0.03$.
We derive a scatter of 0.09 dex and 0.12 dex for the CCF and JAVELIN datasets, 
respectively; each is insensitive to the size of the adopted errorbar (see Appendix~\ref{S:AppendixA}). 
For completeness, we report two types of scatter in Table~\ref{tab7} and Figure~\ref{fig:rl1}: 
one relative to the native $R(\Hbeta) - L$(5100\AA) relationship for NGC\,5548 
(red dashed curve) and the scatter contribution of this source to the global 
relationship (with slope $\beta$= 0.53; black solid curve), which is the scatter of prime interest to this study. 
 We infer a larger amount of scatter based on 
the JAVELIN dataset compared to the CCF dataset.  This is easily understood 
because the measured native slope is steeper in this case and the degree of estimated scatter 
depends\footnote{The sum of the quadratures of the measurement uncertainties and the estimated scatter $\epsilon_0$ 
(see equation\,(\ref{E:rlngc5548})), respectively, must sum to the quadrature of the observed scatter, the $\sigma_{rms}$.} 
 on the amplitudes of the 
uncertainties of the $R_{BLR}$ and $L(5100\AA)$ measurements and the JAVELIN 
dataset has smaller $R_{BLR}$ uncertainties.

Since we demonstrate above that the variability of NGC\,5548 is representative of
reverberation-mapped AGN, 
we can extrapolate these results to predict that the variability of individual
objects will add a scatter of order 0.1\,dex into the global 
$R(\Hbeta)-L$(5100\AA) relationship. We verify this in Appendix~\ref{S:montecarlo} by means
of Monte Carlo simulations.

 \subsubsection{The $R-L$(UV) Relationship of NGC\,5548}\label{S:3.3.2}

\begin{figure} 
\begin{center}$
\begin{array}{c}
\includegraphics[scale=1.0]{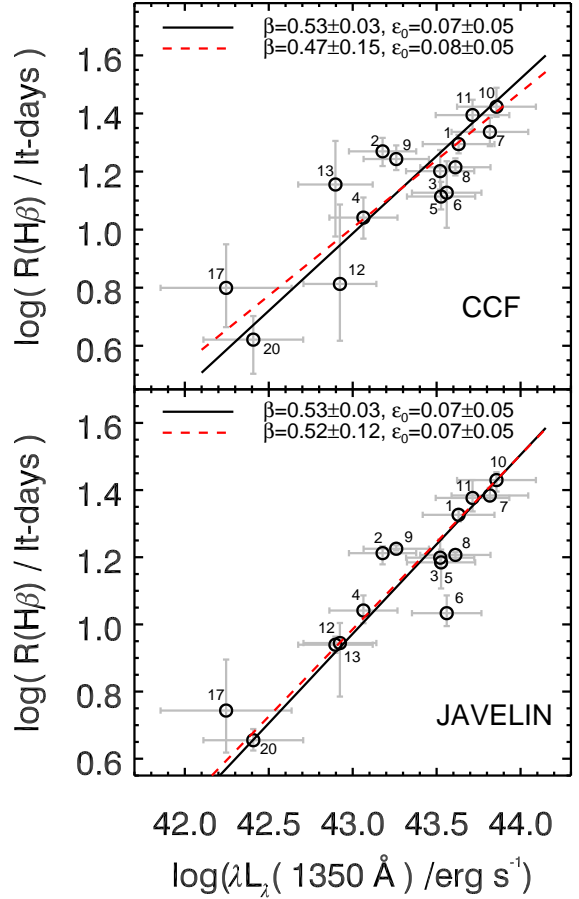}\\ 
\end{array}$
\end{center}
\caption{The $R-L(1350\AA)$ relationships for NGC\,5548 using the CCF dataset (top), 
and the JAVELIN dataset (bottom).  The $L(1350\AA)$ luminosities are computed based on 
the measured $L(5100\AA)$ values and the $L$(optical)$-L$(UV) analysis as outlined in
Section~\ref{S:3.3.2}.
See Figure \ref{fig:rl1} for symbols and color code.}
\label{fig:rluv1}
\end{figure}

Because $L$(ionizing) is the luminosity that sets the BLR size, and $L$(UV) is closer in energy to $L$(ionizing) than $L$(optical), we test here if an inferred native  $R$(\Hbeta) $-$ $L$(1350\AA) relationship for NGC\,5548 will have a slope of $\sim$0.5, more consistent with the physical expectations. This is a zeroth-order test 
because the $L$(1350\AA) values are not direct measurements but inferred
from the available optical luminosities for most $R$(H$\beta$) measurements. 
To convert the $L$(5100\AA) values to $L$(1350\AA), we use the 
$L$(5100\AA) $- L$(UV) relationship established for the `NGC\,5548 All' 
dataset and given in Table~\ref{tab:lopluvlinmix}. 
The data from the two separate monitoring years do produce somewhat
different $L$(5100\AA) $- L$(UV) slopes. However, using all available data
to cover a longer temporal baseline over which to calculate a single 
relationship is likely to be more representative\footnote{
If we adopt the `Year 1' slope for the conversion, the slope changes but
the observed scatter inferred from a similar analysis to that outlined below 
does not change significantly. }
of the overall relationship between $L$(5100\AA) and $L$(UV) over 
the time scales covered by our full set of $L$(5100\AA) measurements.

To convert $L(5100\AA)$ to $L(1350\AA)$ we adopt the following parameterization, 
obtained by regressing the NGC\,5548 data with $L$(5100\AA) as the independent 
measurement\footnote{This is consistent with inverting equation 
(\ref{E:linmixerr}) but setting $\epsilon_i =0$.}: 
%
\begin{equation}
\log \Bigg[\frac{\lambda L_{\lambda}(1350\AA)}{{\rm erg\, s}^{-1}}\Bigg] 
= 43.06\pm0.10 +(1.73\pm 0.34) \log \Bigg[ \frac{\lambda L_{\lambda}(5100\AA)}{10^{43} {\rm erg\, s}^{-1}} \Bigg] . 
\label{E:lopluvN55}
\end{equation}

The normalizations of the luminosities are introduced to better constrain 
the uncertainties in the regression procedure \citep[e.g.,][]{Tremaine2002}.  
The errors on $L$(1350\AA) are 
propagated from the uncertainties in the $L$(5100\AA), slope, and intercept 
values according to standard error propagation rules 
\citep[][]{Taylor} 
and are therefore larger than we expect from direct UV measurements. 
The results of this luminosity conversion are demonstrated in 
Figure~\ref{fig:rluv1}, which shows the $R(\Hbeta) - L$(1350\AA) relationships
based on the CCF (top) and JAVELIN (bottom) datasets.

Regression analysis on these new relationships yields best-fit slopes for the CCF and JAVELIN datasets of $\beta$\,=\,0.47$\pm$0.15 and $\beta$\,=\,0.52$\pm$0.12, respectively,
close to the theoretically expected slope.  
This suggests that $L$(UV) is a better proxy of $L$(ionizing) 
and for that reason we may also expect a reduction in the scatter of the $R(H\beta) - L$ relationship by adopting the $L$(UV) luminosity. Unfortunately, we cannot strictly address this latter issue because the larger propagated uncertainties on $L$(1350\AA) in our current investigation may be suppressing the estimated scatter artificially.

The similar work of \citet{Bentz2007} also obtains best-fit slopes $\sim$0.5. However,
the current work supersedes that earlier effort because it is based on 
(a) a larger database of NGC\,5548 monitoring data with improved flux calibration; 
(b) improved luminosity measurements and uncertainties owing to improved host 
galaxy light determinations, Galactic extinction corrections, and updated 
uncertainty determinations;
(c) an analysis to specifically address how the connection between the 
optical and UV luminosities factors into the scatter in the observed 
global $R(\Hbeta)-L$(5100\AA) relationship based on $R(\Hbeta)$ values derived from 
both the classical CCF method and the new JAVELIN method; 
(d) Monte Carlo simulations (Appendix~\,\ref{S:montecarlo}) to predict the effects of the native  $R(\Hbeta)-L$(5100\AA) and
$R(\Hbeta)-L$(1350\AA) relations on the corresponding global relationships (addressed next).

\section{Discussion}\label{S:Dis}

\subsection{The Scatter in the Global $R-L$ Relationships}\label{S:dis2}

Since the relationship was first established \citep[e.g.,][]{Kaspi2000,Kaspi2005}, 
the largest improvement imposed was the correction for host star light 
contamination of the optical luminosities \citep[e.g.,][]{Bentz2006a, Bentz09a} that
changed the global slope from $\sim$0.7 to 0.54. 
Upon correcting for host galaxy contamination for the 35 AGN in the RM sample at the time,
 \citet{Bentz09a} estimate the observed scatter, $\epsilon$, using the FITEXY method \citep{Press92} to be $\sim$40\%  or 0.15\,dex (in $R_{BLR}$).
As the measurements in recent years have improved, the observed $R(\Hbeta)-L$(5100\AA) 
relationship has become increasingly tighter.  \citet{Peterson2010} found a scatter 
of just 0.11\,dex when including only the most robust measurements, namely those 
based on light curves so well-behaved that the time delay can be estimated by eye. 
The most recent work \citep{Bentz2013} suggests that for 41 nearby AGN, that cover a wide optical luminosity range from $10^{42}$ {\rm erg\, s}$^{-1}$ to $10^{46}$ {\rm erg\, s}$^{-1}$, the observed scatter amounts to 0.19\,dex when all data are included. When restricting the analysis to the better dataset where two AGN, Mrk\,142 and PG\,2130$+$099, with poorly constrained lags\footnote{Recent work suggest that the lag measurements for PG\,2130$+$099 and Mrk\,142 are in error. See \citet{Grier2013}, \citet{Bentz2013},  and \citet{Du2014} for details and discussion.} are omitted, the more robust Bayesian regression method of Kelly (2007) reveals a scatter of 0.13\,dex. 
The \citet{Bentz2013} study includes new measurements of low-luminosity AGN and improved corrections for host galaxy light and Galactic reddening.

Understanding the origin of the observed scatter can help us understand how
to minimize the scatter for future studies and application of the relationships 
and help us understand the underlying physics of the relationships.
While high-quality data and accurate lag measurements are important for application to 
precision measurements of black hole mass and cosmic distances, other 
issues affect the scatter.  \citet{Bentz2013} find that a large contribution 
to the observed scatter is in fact the accuracy to which we know the physical 
distance to some of the nearby AGN. These objects are so nearby that they do 
not follow the Hubble flow but have significant peculiar velocities, making 
their redshifts poorly suited for distance measurements; alternative distance measurements are often lacking or have large errors.
\citet{Watson2011} also discuss known contributions to the currently observed scatter due to uncorrected but significant internal reddening in a few objects and remaining inaccurate lag measurements, the latter of which recent or ongoing studies are continuing to address \citep[see e.g.,][]{Du2014,Peterson2014}.
The main goal of our investigation has been to quantify the amount of scatter
that can be introduced by the use of $L$(5100\AA) 
in the observed global radius $-$ luminosity relationship 
and which can potentially be mitigated by adopting a better proxy of the
ionizing luminosity $L$(ionizing) that drives this relationship (see eqn.\,(\ref{E:ion})),
such as the UV luminosity $L$(1350\AA).

In our study of how $L$(5100\AA) changes with $L$(1350\AA) for a sample 
of six RM AGN (\S~\ref{S:3.1}; Figure~\ref{fig:Lrel}), we find that for each
object the two luminosities are not linearly related (moreover, the global slope is $\sim$0.84; Table~\ref{tab:lopluvlinmix}) 
and therefore not directly interchangeable. Also, the shallow slopes and the $F_{var}$ values (Table\,\ref{tab:fvar}) indicate stronger UV variability, consistent with previous observations 
\citep[e.g.,][]{Clavel91,Korista95,Vandenberk2004,Wilhite2005,Zuo2012}.
Based on our Bayesian analysis, we find that the non-linearity between the optical and UV 
luminosities for individual AGN may introduce a scatter into the observed global
$R(\Hbeta) -L$(5100\AA) relationship of 0.09\,dex (Table~\ref{tab:lopluvlinmix}), which is the best constraint that can be placed on this effect with the currently available data.

To study the impact of long-term variability on the scatter in the global
$R$(\Hbeta) $-$ $L$(5100\AA) relation, we turn to the well-studied  Seyfert 1 galaxy, 
NGC\,5548, the only source for which we have data spanning decades. 
Although this is just a single source, its variability nature is 
representative of the current RM sample (\S\,\ref{S:2.2}) and it is fair to assume that 
this AGN will provide a representative measure of the observed scatter in the global
$R(\Hbeta)-L$(5100\AA) relationship. 

While we analyze both the CCF and JAVELIN datasets of $R(\Hbeta)$ values in 
\S~\ref{S:3.3} we mainly focus on the results based on the standard CCF method
as this allows a direct comparison with previous work.
On account of the steep slope ($\beta$ = 0.79$\pm$0.20; Table~\ref{tab7}) and 
the scatter ($\epsilon_0$ = 0.08\,dex) observed for the native 
$R(\Hbeta)-L$(5100\AA) relationship of NGC\,5548, traced by its intrinsic variability,
our Monte Carlo simulations (Appendix\,\ref{S:montecarlo}) show that if this native  relationship is
representative of each of the 39 AGN in the current global $R$(H$\beta$) $- L$(5100\AA) 
relationship, we can expect a scatter of 0.11\,dex $\pm$0.01\,dex in the global relationship,  
a significant fraction of the current scatter of 0.13\,dex measured by \citet{Bentz2013}.
 This estimate may likely be an upper limit to the scatter we can expect in the global
relationship, because most AGN and, especially, the higher luminosity quasars are expected 
to vary with smaller luminosity amplitudes than assumed in our simulations, as discussed in Appendix\,\ref{S:montecarlo}. 
We find that for a decreasing typical variability amplitude for the AGN on the global relationship, the global scatter approaches a floor just above the level of the assumed native  scatter, $\epsilon_0$. This emphasizes the importance of the scatter in the relationships traced by individual AGN, as they vary intrinsically, for the global scatter.
Future work should focus on better understanding this scatter, as this is outside the scope of this work. 
The fact that the steep single-object slope alone can account for $\sim$0.08\,dex 
of the expected 0.11\,dex global scatter implies that a significant fraction of the estimated scatter of 0.13\,dex measured for the empirical global relation
\citep{Bentz2013} can potentially be mitigated by adopting a more accurate proxy for $L$(ionizing).
In that case, we expect the native, single-object  slope to be close to 0.5, as we see for the inferred native  $R$(\Hbeta) $-$ $L$(1350\AA) relationship for NGC\,5548 (\S\,\ref{S:3.3.2}), such that the individual AGN vary {\it along} the global relationship.
This particular situation highlights, again, the importance of the scatter in the relationship of individual AGN, which propagates directly through to the global relationship.

\paragraph{Effects of JAVELIN-based measurements.}
We note that our parallel analyses of the $R(H\beta) - L$(5100\AA) and 
$R(H\beta) - L$(1350\AA) relationships based on $R(\Hbeta)$ measurements with the 
JAVELIN analysis method corroborate the regression results based on the 
standard CCF method, showing a steeper slope of the optical relation and a slope
consistent with 0.5 for the relation based on UV luminosities. 
However, they also suggest that the actual  scatter in these relationships 
may be larger than estimated previously by means of the CCF method \citep[e.g.,][]{Bentz2013}. 
The estimated scatter depends on the measurement uncertainties (i.e., the size of the error bars). 
With its fuller use of information, the JAVELIN method yields $R(\Hbeta)$ measurements 
with smaller measurement errors. As a result, we infer a larger scatter for the JAVELIN-based 
radius $-$ luminosity relationships. Specifically, we estimate a 
scatter of 0.12\,dex $\pm$\,0.04\,dex for the $R(H\beta) - L$(5100\AA) relationship of NGC\,5548,
$\sim$30\% larger than the value inferred based on the classical CCF method.
Repeating the Monte Carlo simulations of Appendix\,\ref{S:montecarlo} for the native  
$R(H\beta) - L$(5100\AA) relationship for the JAVELIN method, we obtain a predicted mean scatter
$\mu_{\rm mock}$ = 0.13\,dex$\pm$0.01\,dex for the RM sample, assuming all the AGN vary like NGC\,5548.
However, we cannot quantify the relative contribution of this scatter to the scatter in the global relationship because, at present, JAVELIN-based lags \citep{Zu2011} do not exist for the full dataset presented by \citet{Bentz2013}.

\subsection{On the slope differences between the native  and global relationships.}\label{S:dis3}
Given the photoionization physics predictions that $R \propto L({\rm ionizing})^{0.5}$,
it is notable that when we use $L$(5100\AA), as opposed to $L$(ionizing), in the observed 
radius $-$ luminosity relationship the global slope is very close to a value of 0.5
\citep{Bentz09a, Bentz2013}. 
Yet, on the contrary, the native, single-object $R(\Hbeta)-L$(5100\AA) relationship appear to be 
somewhat steeper than this global relation. 
What may appear as a conundrum is in fact easily explained by a general AGN property
and the intrinsic source variability properties studied here. 
The regression results in Table~\ref{tab:lopluvlinmix} show that the mean optical and UV luminosities of individual AGN, marked in Figure~\ref{fig:Lrel} by black symbols, trace a linear $L$(optical) $-$ $L$(UV) relationship across the AGN sample to within the uncertainties: the mean luminosities scale with a power of $0.96\pm 0.21$. This can be understood from the perspective that a more massive black hole will result in a higher mean luminosity that on average scales equally across the optical-UV region. This means that the average optical luminosity is typically a good proxy of the average UV luminosity. If the UV luminosity is a good proxy of $L$(ionizing), so is the mean optical luminosity and we can expect a global slope of the $R(\Hbeta) - L$(5100\AA) relationship close to 0.5.
The steep slope of the native  $R(\Hbeta)-L$(5100\AA) relationship for individual objects, as seen for
NGC\,5548 in Figure~\ref{fig:rl1}, is simply a result of the optical source
flux varying typically with smaller amplitudes than the UV flux. This is verified by the 
higher $F_{var}$ values for the UV continuum (Table~\ref{tab:fvar}) and the
shallow slopes of the $L$(optical) $-$ $L$(UV) relationship for individual AGN (Figure~\ref{fig:Lrel}).
That this is a luminosity color effect is confirmed by the native  $R(\Hbeta)-L$(1350\AA) relationship for NGC\,5548
having a slope\,$\sim$0.5.

\subsection{Alternate Proxy for the Ionizing Luminosity}

\begin{figure} 
\begin{center}$
\begin{array}{c}
\includegraphics[scale=0.6]{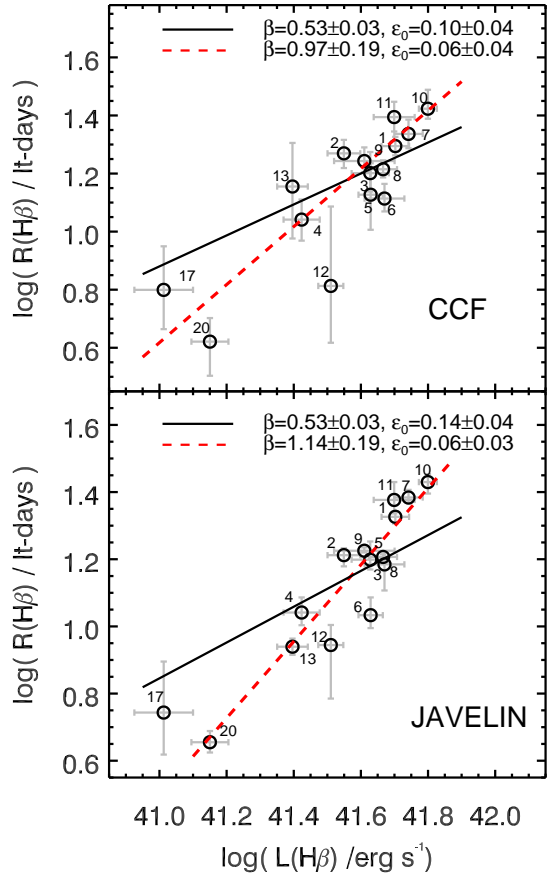}\\ 
\end{array}$
\end{center}
\vskip -0.3cm
\caption{The $R-L(\Hbeta)$ relationships for NGC\,5548 using the CCF dataset (top), 
and the JAVELIN dataset (bottom). The $L(\Hbeta)$ luminosities (Table~\ref{tab:ngc5548data}) are measured from the recalibrated mean spectra (see \S\,\ref{S:2.2} for details).
See Figure \ref{fig:rl1} for symbols and color code.}
\label{fig:fig5}
\end{figure}

The H$\beta$ line luminosity, $L(\Hbeta)$, is considered to be a good measure of 
the ionizing luminosity because \hbeta{} is a recombination line \citep{Osterbrock2006} 
and it, therefore, carries the potential of providing a good readily accessible 
proxy in the optical observing window \citep{Wu2004, Greene2010}.  For that reason, one might expect that 
$L$(H$\beta$) could provide a better measure of slope and scatter of the 
intrinsic $R(\Hbeta) - L$(ionizing) relationship than the propagated $L$(1350\AA). 
Unfortunately, the $R(H\beta) - L(H\beta)$ relationship does not offer any more information 
than that of the UV luminosities estimated in this work.
Because H$\beta$ exhibits a Baldwin Effect when time delays are correctly accounted for 
\citep{Gilbert2003}, there is not a one-to-one correspondence between
the number of ionizing photons and the number of H$\beta$ photons. 
Instead, the line equivalent width decreases with increasing ionizing flux because 
the line responsivity (i.e., the efficiency by which ionizing photons are converted to line photons) 
becomes less efficient \citep{Korista2004}. 
We verified this by measuring the narrow-line subtracted H$\beta$ luminosity 
listed in Table\,\ref{tab:ngc5548data})  from the recalibrated mean spectra 
of NGC\,5548 obtained from each of the epochs spanning the 20-year monitoring 
database available (\S\,\ref{S:2.2}).  The resulting $R(H\beta) - L(H\beta)$ relationships 
(Figure~\ref{fig:fig5}) exhibit an even steeper slope than that of $R(H\beta) - L$(5100\AA ) 
and only by applying a similar correction for the luminosity color \citep{Gilbert2003} 
as that applied here, do we confirm that the $R(\Hbeta) - L$(UV) relationship has a 
slope of 0.5.
We therefore conclude that the best proxy for the ionizing luminosity is a directly
measured luminosity at energies close to the peak of the ionizing spectral energy 
distribution, such as $L$(UV).

\subsection{Implications for Cosmology Studies}

The amount of observed scatter in the global $R(\Hbeta)-L$(5100\AA) relationship is important for cosmological implications \citep{Watson2011}. 
It introduces an uncertainty in the inferred luminosity for a given measured
$R(\Hbeta)$ and, as a result, in the luminosity distance.  However, to use the relationship as a distance indicator, it is reasonable to use the better data and exclude clearly bad measurements.
\citet{Bentz2013} show that the current $R(\Hbeta) - L$(5100\AA) relationship has an 
observed scatter of 0.13\,dex when Mrk\,142 and  PG\,2130$+$099 with a poorly constrained lags are omitted, which corresponds to an uncertainty in the distance modulus of $\Delta \mu$\,=\,0.33\,mag.
This is already an improvement over the value of 0.5\,mag reported by \citet{Watson2011}.
Our analysis shows that by adopting a more accurate proxy of $L$(ionizing) than $L$(5100\AA), such as $L$(UV),
we may eliminate a scatter of up to $\sim$0.08\,dex (as estimated in this work based on the CCF method),
thereby bringing the total observed  
scatter of 0.13\,dex to 0.10\,dex and reduce the uncertainty  to $\Delta \mu$\,=\,0.26\,mag. 
The scatter in the global $R$ $- L$(UV) relationship depends, however, on the scatter in the native relationships for individual objects as they vary intrinsically. This scatter may well be lower for a better proxy of $L$(ionizing). 
With additional attention to other sources of uncertainties and scatter in the
$R$(\Hbeta) $-$ $L$(ionizing) relationship, such as reddening, improved $R(\Hbeta)$ lag 
measurements, and distance measurements for some of the most nearby RM AGN 
\citep[see][for discussion]{Watson2011, Haas2011, Bentz2013}, the observed scatter 
can potentially be reduced further. For the $R$(\civ) $- L$(1350\AA) relationship
applied to high-redshift AGN, our discussion earlier emphasizes again the importance of 
understanding the origin of the scatter in the native  relationship of each individual AGN 
as the typical 
scatter in the single-object relationships defines the scatter in the global relationship, in this case.
Future studies of large reverberation mapping datasets, obtained from multi-object spectroscopic monitoring campaigns of hundreds of AGN (some of which are currently underway), hold promise to establish 
how this scatter in the global relationships can be mitigated or minimized through a better understanding of the potential systematics involved.
As a result, there is a large potential for the global $R$(\civ) $- L$(1350\AA) relationship to be 
a competitive luminosity distance indicator, both at low and high redshift.

\section{Conclusion}\label{S:Con}

Since the ionizing luminosity is what drives the radius $-$ luminosity
relationship, we have investigated whether the use of the optical 
luminosity $L$(5100\AA), as opposed to the ionizing luminosity, can account 
for some of the scatter in the observed global $R(\Hbeta)-L$(5100\AA) relationship
\citep[e.g.,][]{Bentz2013}. 
Based on our analysis of the relationship between multiple near-simultaneous
pairs of optical and UV continuum luminosity measurements (to within two days)
available for six reverberation-mapped AGN (NGC\,5548, NGC\,7469, NGC\,3783, 
NGC\,4151, 3C\,390.3, Fairall\,9), the long-term optical and UV 
continuum flux variations of Seyfert 1 galaxy, NGC\,5548, and a suite of Monte
Carlo simulations, our main findings are as follows:

\begin{enumerate}
\item
We present the most recent updates of the native  $R(\Hbeta) - L$(5100\AA) relationship for NGC\,5548, traced by its intrinsic variability that takes into account the recalibration of the flux measurements of \citet{Peterson13} and the updates of \citet{Bentz2013}. We present the relation for the $\Hbeta$ lags, $R(\Hbeta)$, determined by both the CCF and the JAVELIN methods (Table\,\ref{tab7}), finding slightly steeper slopes $\beta$ = 0.79  and $\beta$ = 0.88, respectively, than previously reported. The scatter measured in this native  relation amounts to 0.07\,dex $-$ 0.08\,dex.  We also present JAVELIN-based lags of the Year 17 \citep{Bentz2007} and  Year 20 \citep{Bentz09b} monitoring campaigns, not included in the \citet{Zu2011} study.

\item We confirm $L$(1350\AA) to be a better proxy for $L$(ionizing) than is $L$(5100\AA).
Our analysis of the native  NGC\,5548 $R(\Hbeta) - L$(1350\AA) relationship shows a slope consistent 
with the theoretically expected slope $\beta$=0.5 in contrast to the native  $R(\Hbeta) - L$(5100\AA) relation.

\item
The $\Hbeta$ luminosity is not a more suitable substitute for the ionizing
luminosity than $L$(5100\AA) as it needs a similar color correction.

\item 
The typical lower variability amplitudes of the AGN optical continuum compared to the UV continuum suggest that the native $R(\Hbeta) - L$(5100\AA) relationship for individual AGN will typically be steep, as seen for NGC\,5548.  
If all AGN vary like NGC\,5548 with a similar slope of their individual relationship, this steep slope alone will contribute a typical scatter of 0.08\,$\pm$\,0.01\,dex (Appendix\,\ref{S:montecarlo}) 
to the currently observed scatter of 0.13\,dex \citep{Bentz2013} in the global $R(\Hbeta) - L$(5100\AA) relationship. This suggests that a sizable fraction of the observed scatter can be mitigated by the use of a UV luminosity in lieu of $L$(5100\AA).

\item
Assuming NGC\,5548 is representative of the AGN population, the combined effect of the steep slopes and the scatter in the single-object $R(\Hbeta) - L$(5100\AA) relationships can account for most ($\sim$0.11\,dex) of the current scatter  in the observed global $R(\Hbeta) - L$(5100\AA) relationship.

\item
A significant contribution to the scatter in the global $R$(\Hbeta) $- L$(5100\AA), $R$(\Hbeta) $- L$(1350\AA), and $R$(\civ) $- L$(1350\AA) relationships comes from the scatter in the corresponding relationships traced by the individual AGN as they exhibit intrinsic luminosity variations. If the native $R$(\Hbeta) $- L$(5100\AA) relationship for NGC\,5548 is typical for AGN, then it can contribute a scatter $\sim$0.08\,dex to the global scatter, which is about half of the current observed scatter.
To minimize the global scatter we need to better understand this scatter in the relationships for individual AGN. Future studies will need to focus on this effort as it is beyond the scope of the current work.

\item
By adopting a UV luminosity as a better proxy for the ionizing luminosity than $L$(5100\AA),
the scatter in the global $R(\Hbeta) -L$(UV) relationship is expected to be lower by $\sim$0.08\,dex
than the current global $R$(\Hbeta) $- L$(5100\AA) relationship.
This is expected to invoke a reduction of the uncertainty in the distance modulus from 0.33\,mag to
0.26\,mag for cosmic distances derived from the $R(\Hbeta) -L$(UV) relationship.
A further decrease of this uncertainty is expected when the scatter in the relationships traced by the intrinsic variability of individual AGN is better understood and when object-to-object differences, such as internal reddening, are corrected for.

\item
Even though we see a steeper $R$(\Hbeta) $- L$(5100\AA) relationship for individual AGN
than the global relationship for the entire reverberation mapped sample $-$ because AGN typically vary with lower optical luminosity amplitudes than the ionizing luminosity that drives the relationship $-$ the average optical luminosity of a given AGN is an equally good proxy of the average ionizing luminosity as the UV luminosity (\S\,\ref{S:dis3}).

\end{enumerate}

By extrapolating these result we can expect 
that a well-populated version of the existing, but tentative, $R$(\civ)$-L$(1350\AA) 
relationship \citep{Kaspi07} will similarly have less observed scatter.
Along with the emphasis made by \citet{Bentz2013} on the pressing 
need to obtain accurate distances of the nearest AGN that define the lower end of this relationship, 
there is therefore a strong impetus to obtain additional monitoring data in the restframe UV energy range.

\subsection*{Acknowledgments}  
We thank Michael Goad, Kirk Korista, and Roberto Assef for helpful discussions.  
Brandon Kelly is thanked for helping with minor modifications of the 
`$linmix\_err$' code to allow estimates of the scatter for a given, 
fixed slope.  This work has benefited from the discussions at the workshop 
on {\it Improving Black Hole Mass Estimates in Active Galaxies} at the DARK Cosmology 
Center, 2012 July Copenhagen, Denmark. MV acknowledges support from a
FREJA Fellowship granted by the Dean of the Faculty of Natural Sciences at the
University of Copenhagen and a Marie Curie International Incoming Fellowship. 
KD acknowledges support from a Marie Curie International Incoming Fellowship
and a National Science Foundation Postdoctoral Fellowship.
MV thanks the Kavli Institute for Theoretical Physics at University of California,
Santa Barbara for their hospitality while finalizing this work.
The research leading to these results has received funding from the People 
Programme (Marie Curie Actions) of the European Union's Seventh Framework 
Programme FP7/2007-2013/ under REA grant agreement No. 300553 (MV and KD).
The Dark Cosmology Centre is funded by the Danish National Research Foundation. 
This research was supported in part by the National Science Foundation 
through grant AST-1008882 to The Ohio State University (BMP), grant AST-1302093 (KD),
CAREER Grant AST-1253702 to Georgia State University (MCB), 
and under Grant No. NSF PHY11-25915 (MV).
This research has made use of the NASA/IPAC Extragalactic Database (NED), which 
is operated by the Jet Propulsion Laboratory, California Institute of Technology, 
under contract with the National Aeronautics and Space Administration.

\appendix
\section{A. Error-bar Sensitivity Test}\label{S:AppendixA}
Neither the Bayesian regression method 
\citep{Kelly2007} nor the FITEXY method of \citep{Press92}
can account for the asymmetric uncertainties in our $R_{BLR}$ measurements 
(\S~\ref{S:3.3.1}). 
Therefore, we performed a so-called `error-bar sensitivity test' to test 
how sensitive the regression analysis is to the adopted (symmetric) error bar.
In this test, for the symmetric measurement errors in the $R_{BLR}$ values we 
assume either: 
\begin{description}
\vskip -0.4cm 
\item[a)]
 the `positive' $1\sigma$ uncertainties (i.e., upper error bar), 
\item[b)]
 the `negative' $1\sigma$ uncertainties (i.e., lower error bar), 
\item[c)]
 the error computed as: $([\sigma{\rm (positive)}^2 + \sigma{\rm (negative)}^2]/2)^{1/2}$, 
since this behaves correctly in the limit $\sigma$(positive) = $\sigma$(negative), 
\item[d)]
 the `largest' of the two $1\sigma$ uncertainties, 
\item[e)] the `smallest' of the two $1\sigma$ uncertainties, or 
\item[f)] the error bar that points toward the fitted relation.  
\end{description}

\vskip -0.2cm 
Option `f' involves an iterative process after an initial selection of error 
bars until the relative change in slope and zero point between two iterations is 
less than $10^{-11}$; typically, only about four iterations are needed.
To test the sensitivity of option `f' to the choice of error-bar in the first 
iteration, we run this test three times: first with the `positive' uncertainty 
(option a), second with the `negative' uncertainty (option b), and third with 
the `largest' uncertainty (option c).

Applying this `error-bar sensitivity test' to the $R(\Hbeta) - L$(5100\AA) 
relationship of NGC\,5548 we find slopes in the range from 0.79 to 0.85 with 
similar uncertainties of $\pm$0.20 for the CCF dataset.  For the JAVELIN dataset  
we find slopes between 0.88 to 0.90 with uncertainties of $\pm$0.17. 
The results are insensitive to the choice of the initial error-bar in option `f'.  
For example, the difference in slope (and uncertainty) is at most 0.01 when the extreme 
`largest' and `smallest' error bars are adopted. 
The change of the slope given the adopted error bar is in any case within 
the $1\sigma$ uncertainty.  
 
Our estimates of the scatter in the $R(\Hbeta) - L$(5100\AA) relationship of NGC\,5548 
for the CCF data are 
0.075$\pm$0.040 dex and 0.077$\pm$0.043 dex when we adopt the `largest' 
and `smallest' uncertainties, respectively. 
For the JAVELIN dataset the equivalent values are 0.070$\pm$0.039 dex and 
0.073$\pm$0.037 dex, respectively. 
Thus, in both datasets, the inferred scatter is essentially 
the same for these two extreme error-bar settings.  Similarly, 
we find the scatter to be insensitive to the initial choice 
of error-bar (option `f') $-$ it agrees to within the 
1\,$\sigma$ uncertainty of $\sim$0.04 dex.
 
When we apply this sensitivity test to the native  $R(\Hbeta) - L$(1350\AA) 
relationship of NGC\,5548 we see no or insignificant changes in the slopes.
We find best fit slopes of 0.47\,$\ltsim$$\beta$(CCF)\,$\ltsim$0.50 and 
0.51$\ltsim$\,$\beta$(JAVELIN)\,$\ltsim$0.53 with uncertainties of 
$\sim$0.15 (CCF) and  $\sim$0.12 (JAVELIN).  
The slopes based on the `largest' and `smallest' uncertainties are 
$0.47\pm0.15$ and $0.49\pm0.15$, respectively, for the CCF dataset. 
The equivalent values for the JAVELIN dataset are $0.52\pm0.12$ and $0.52\pm0.15$, 
respectively. 
For the different sub-options of option `f' there is no difference in the slopes. 
We determine the  scatter in the UV relationship to 
be in the range of 0.078 dex$ - $0.083 dex  and 0.067 dex$ - $0.072 dex 
for each of the CCF and JAVELIN datasets, respectively.  

In summary, we have demonstrated very little sensitivity of the slopes 
and the estimated scatter to the specific adopted symmetric error-bar 
for either of the datasets and for either of the radius $-$ luminosity
relations addressed here. 
While the differences are insignificant, we are conservative and adopt the 
`largest' 1\,$\sigma$ uncertainty for our regressions throughout (Table~\ref{tab7}).

\section{B. The Effect of Intrinsic AGN Variability on the Global $R$(\hbeta) $- L$(5100\AA) Relationship: Monte Carlo Simulations}
\label{S:montecarlo}

Our analysis of the native  $R(\Hbeta) - L$(5100\AA) relationship for NGC\,5548 in \S~\ref{S:3.3.1}
shows that its steep slope can introduce a scatter of order 0.1\,dex in the global 
$R(\Hbeta) - L$(5100\AA) relation.
 If this steep native  relationship is characteristic for all the reverberation
mapped AGN, the question remains: how much of the current scatter in the global $R(\Hbeta) - L$(5100\AA)
relation is due to this effect? 
To examine this, we performed Monte Carlo simulations using mock databases of $R$ and $L$ pairs that sample
the native  NGC\,5548 $R(\Hbeta) - L$(5100\AA) relationship and apply it to
the sample of reverberation mapped AGN presented by \citet{Bentz2013}.

To generate the mock database, we assume that each AGN in the sample varies in a similar manner to NGC\,5548.
We let each object vary in luminosity along its own native  $R$(\Hbeta)\,$-$\,$L$(5100\AA) relationship with an assumed slope $\beta$\,=\,0.79 and scatter $\epsilon_{0}$\,=\,0.081\,dex in $R(\Hbeta)$ (i.e., applying the results of the CCF dataset shown in the top panel of Figure~\ref{fig:rl1}). We describe 
the luminosity distribution by a Gaussian function centered at the mean luminosity, $<$\,$L$\,$>$, observed for each AGN \citep{Bentz2013} with a standard deviation of $\sigma_L$\,=\,0.30\,dex, as measured for NGC\,5548 for the 15 epochs analyzed here (\S\,\ref{S:2.2}).
For the purpose of this test, a Gaussian function is a reasonable approximation of AGN variability behavior, which can be described 
by a damped random walk \citep[e.g.,][]{Kelly2009, Kozlowski2010}, i.e., a stochastic process with an exponential covariance matrix.
We perform 2000 Monte Carlo realizations, where each realization samples a given object at a random point along its simulated native  relationship, and we make a mock global $R$(\Hbeta)\,$-$\,$L$(5100\AA) relationship by sampling 69 ($R$, $L$) pairs for the 39 AGN to match the sample size presented by \citet{Bentz2013}; i.e., we use multiple ($R$, $L$) pairs for objects\footnote{However, our tests show that the results are insensitive to how the 69 data pairs are selected among the AGN, as expected given the assumed similar variability properties of all the objects.}  where \citet{Bentz2013} includes multiple RM results. 
Next, for a given realization of the mock global $R(\Hbeta) -L(5100\AA)$ relationship, we compute for each of the 69 randomly selected ($R$, $L$) pairs, the residuals between the simulated values of $R$ and the radius predicted from the \citet{Bentz2013} $R(\Hbeta) -L(5100\AA)$ relationship. For the resulting distribution of the 69 $R$-residuals, we adopt the standard deviation as the observed scatter for this particular realization.

With 2000 realizations we obtain a distribution of the estimated scatter  
with a mean (i.e., the most likely scatter) and standard deviation of $\mu_{MOCK}\pm\sigma_{MOCK}$\,=\,0.106\,$\pm$\,0.009 dex. 
This means that the combination of a steep native  relationship and an assumed scatter in $R$ around the native  relationship of $\epsilon_0$\,$\sim$\,0.08\,dex will result in a typical scatter in the global relationship of $\sim$0.11\,dex, if all AGN vary like NGC\,5548. 
The steep slope of the native  relation alone (i.e., when $\epsilon_0$=0) contributes a mean scatter of $\mu_{MOCK}$\,= \,0.077\,dex\,$\pm$\,0.007\,dex in the global relationship, entirely consistent with the scatter contributions adding in quadrature.

In reality, each AGN will vary with a different native slope, $\beta$, and scatter, $\epsilon_0$, and a different $\sigma_L$ around its mean luminosity, $<$\,$L$\,$>$, and will thereby contribute with a higher or lower scatter to the global $R$(\Hbeta) $-$ $L$(5100\AA) relationship than we estimate for NGC\,5548. 
There are insufficient data to address the likely distributions of $\beta$ and $\epsilon_0$ for the $R$(\Hbeta) $-$ $L$(5100\AA) relationships of individual AGN, but there are indications that $\sigma_L$ is luminosity dependent and therefore lower for quasars \citep[e.g.][]{Vandenberk2004, Bauer2009, Schmidt2010}.
Although Figure\,\ref{fig:hts} shows that NGC\,5548 does not have the most extreme optical variability properties of the RM AGN within a single RM campaign, it has exhibited the largest $L$(5100\AA) differences, $\Delta L$, on long time scales \citep{Bentz2013}; a larger $\sigma_L$ will result in larger $\Delta L$.  
Therefore, by adopting the same  $\sigma_L$\,=\,0.3\,dex value for all AGN in the simulations, we obtain an upper limit on the expected global scatter. Our tests show that as the luminosity distributions narrow to $\sigma_L$ =\ 0.1\,dex, the mean scatter $\mu_{MOCK}$ approaches 0.085\,dex, a scatter slightly higher than the assumed scatter in the native  relationship, $\epsilon_0$. Our analyses also show that as $\beta$ approaches the slope of the global relationship (i.e., the AGN vary {\it along} the global relationship, as opposed to across it), the expected global scatter $\mu_{MOCK}$ converges on the assumed value of $\epsilon_0$. This underscores the importance of further understanding the existing scatter in the single-object relationship and its effect on the global relationship.

Our estimates of the expected global scatter of $\sim$0.11\,dex, assuming all AGN vary like NGC\,5548, is a significant fraction of the observed scatter of 0.13\,dex determined by \citet{Bentz2013} for the empirical global $R(\Hbeta) -L(5100\AA)$ relationship. 
We note that while our simulated native relationships do contain a scatter (or noise) contribution (i.e., $\epsilon_0$\,$>$\,0) that can include some measurement uncertainties due to, for example, small inaccuracies in the $R$ measurements, the scatter we simulated will not take into account other known sources of scatter in the global $R(\Hbeta) -L(5100\AA)$ relationship due to distance measurements or poorly constrained lags that are included in the scatter measured by Bentz et al. (2013).
The fact that we can account for a large fraction of the observed global scatter is good news. Understanding the origin of the scatter means that there is a potential for mitigating it.


\newpage
\newpage
\bibliographystyle{apj}

\begin{thebibliography}{114}
\expandafter\ifx\csname natexlab\endcsname\relax\def\natexlab#1{#1}\fi

\bibitem[{{Baldwin} {et~al.}(1995){Baldwin}, {Ferland}, {Korista}, \&
  {Verner}}]{Baldwin1995}
{Baldwin}, J., {Ferland}, G., {Korista}, K., \& {Verner}, D. 1995, \apjl, 455,
  L119

\bibitem[{{Baldwin}(1977)}]{Baldwin1977}
{Baldwin}, J.~A. 1977, \apj, 214, 679

\bibitem[{{Bauer} {et~al.}(2009)}]{Bauer2009}
{Bauer}, A., {et~al.} 2009, \apj, 696, 1241

\bibitem[{{Bentz} {et~al.}(2009{\natexlab{a}}){Bentz}, {Peterson}, {Netzer},
  {Pogge}, \& {Vestergaard}}]{Bentz09a}
{Bentz}, M.~C., {Peterson}, B.~M., {Netzer}, H., {Pogge}, R.~W., \&
  {Vestergaard}, M. 2009{\natexlab{a}}, \apj, 697, 160

\bibitem[{{Bentz} {et~al.}(2006{\natexlab{a}}){Bentz}, {Peterson}, {Pogge},
  {Vestergaard}, \& {Onken}}]{Bentz2006a}
{Bentz}, M.~C., {Peterson}, B.~M., {Pogge}, R.~W., {Vestergaard}, M., \&
  {Onken}, C.~A. 2006{\natexlab{a}}, \apj, 644, 133

\bibitem[{{Bentz} {et~al.}(2006{\natexlab{b}})}]{Bentz2006}
{Bentz}, M.~C., {et~al.} 2006{\natexlab{b}}, \apj, 651, 775

\bibitem[{{Bentz} {et~al.}(2007)}]{Bentz2007}
---. 2007, \apj, 662, 205

\bibitem[{{Bentz} {et~al.}(2009{\natexlab{b}})}]{Bentz09b}
---. 2009{\natexlab{b}}, \apj, 705, 199

\bibitem[{{Bentz} {et~al.}(2010)}]{Bentz2010}
---. 2010, \apj, 716, 993

\bibitem[{{Bentz} {et~al.}(2013)}]{Bentz2013}
---. 2013, \apj, 767, 149

\bibitem[{{Blandford} \& {McKee}(1982)}]{Blandford82}
{Blandford}, R.~D., \& {McKee}, C.~F. 1982, \apj, 255, 419

\bibitem[{{Cackett} {et~al.}(2007){Cackett}, {Horne}, \&
  {Winkler}}]{Cackett2007}
{Cackett}, E.~M., {Horne}, K., \& {Winkler}, H. 2007, \mnras, 380, 669

\bibitem[{{Cardelli} {et~al.}(1989){Cardelli}, {Clayton}, \&
  {Mathis}}]{Cardelli89}
{Cardelli}, J.~A., {Clayton}, G.~C., \& {Mathis}, J.~S. 1989, \apj, 345, 245

\bibitem[{{Clavel} {et~al.}(1991)}]{Clavel91}
{Clavel}, J., {et~al.} 1991, \apj, 366, 64

\bibitem[{{Collier} {et~al.}(1999){Collier}, {Horne}, {Wanders}, \&
  {Peterson}}]{Collier1999}
{Collier}, S., {Horne}, K., {Wanders}, I., \& {Peterson}, B.~M. 1999, \mnras,
  302, L24

\bibitem[{{Collier} {et~al.}(1998)}]{Collier98}
{Collier}, S.~J., {et~al.} 1998, \apj, 500, 162

\bibitem[{{Crenshaw} {et~al.}(2001){Crenshaw}, {Kraemer}, {Bruhweiler}, \&
  {Ruiz}}]{Crenshaw2001}
{Crenshaw}, D.~M., {Kraemer}, S.~B., {Bruhweiler}, F.~C., \& {Ruiz}, J.~R.
  2001, \apj, 555, 633

\bibitem[{{Crenshaw} {et~al.}(1996)}]{Crenshaw96}
{Crenshaw}, D.~M., {et~al.} 1996, \apj, 470, 322

\bibitem[{{Croton} {et~al.}(2006){Croton}, {Springel}, {White}, {De Lucia},
  {Frenk}, {Gao}, {Jenkins}, {Kauffmann}, {Navarro}, \& {Yoshida}}]{Croton2006}
{Croton}, D.~J., {Springel}, V., {White}, S.~D.~M., {De Lucia}, G., {Frenk},
  C.~S., {Gao}, L., {Jenkins}, A., {Kauffmann}, G., {Navarro}, J.~F., \&
  {Yoshida}, N. 2006, \mnras, 365, 11

\bibitem[{{Davidson}(1972)}]{Davidson1972}
{Davidson}, K. 1972, \apj, 171, 213

\bibitem[{{Davidson} \& {Netzer}(1979)}]{DavidsonNetzer79}
{Davidson}, K., \& {Netzer}, H. 1979, Reviews of Modern Physics, 51, 715

\bibitem[{{De Rosa} {et~al.}(2014)}]{DeRosa2014}
{De Rosa}, G., {et~al.} 2014, \apj, 790, 145

\bibitem[{{Denney} {et~al.}(2006)}]{Denney2006}
{Denney}, K.~D., {et~al.} 2006, \apj, 653, 152

\bibitem[{{Denney} {et~al.}(2010)}]{Denney10}
---. 2010, \apj, 721, 715

\bibitem[{{D{\'{\i}}az-Santos} {et~al.}(2007){D{\'{\i}}az-Santos},
  {Alonso-Herrero}, {Colina}, {Ryder}, \& {Knapen}}]{DiazSantos2007}
{D{\'{\i}}az-Santos}, T., {Alonso-Herrero}, A., {Colina}, L., {Ryder}, S.~D.,
  \& {Knapen}, J.~H. 2007, \apj, 661, 149

\bibitem[{{Dietrich} {et~al.}(1998)}]{Dietrich98}
{Dietrich}, M., {et~al.} 1998, \apjs, 115, 185

\bibitem[{{Dietrich} {et~al.}(2002)}]{Dietrich2002}
---. 2002, \apj, 581, 912

\bibitem[{{Du} {et~al.}(2014)}]{Du2014}
{Du}, P., {et~al.} 2014, \apj, 782, 45

\bibitem[{{Elvis} \& {Karovska}(2002)}]{Elvis2002}
{Elvis}, M., \& {Karovska}, M. 2002, \apjl, 581, L67

\bibitem[{{Fan} {et~al.}(2001)}]{Fan2001a}
{Fan}, X., {et~al.} 2001, \aj, 121, 54

\bibitem[{{Ferrarese} \& {Ford}(2005)}]{FerrareseFord2005}
{Ferrarese}, L., \& {Ford}, H. 2005, \ssr, 116, 523

\bibitem[{{Fynbo} {et~al.}(2013)}]{Fynbo2013}
{Fynbo}, J.~P.~U., {et~al.} 2013, \mnras, 436, 361

\bibitem[{{Gilbert} \& {Peterson}(2003)}]{Gilbert2003}
{Gilbert}, K.~M., \& {Peterson}, B.~M. 2003, \apj, 587, 123

\bibitem[{{Greene} {et~al.}(2010)}]{Greene2010}
{Greene}, J.~E., {et~al.} 2010, \apj, 723, 409

\bibitem[{{Grier} {et~al.}(2012{\natexlab{a}})}]{Grier2012}
{Grier}, C.~J., {et~al.} 2012{\natexlab{a}}, \apjl, 744, L4

\bibitem[{{Grier} {et~al.}(2012{\natexlab{b}})}]{Grier2012b}
---. 2012{\natexlab{b}}, \apj, 755, 60

\bibitem[{{Grier} {et~al.}(2013)}]{Grier2013}
---. 2013, \apj, 764, 47

\bibitem[{{Haas} {et~al.}(2011){Haas}, {Chini}, {Ramolla}, {Pozo Nu{\~n}ez},
  {Westhues}, {Watermann}, {Hoffmeister}, \& {Murphy}}]{Haas2011}
{Haas}, M., {Chini}, R., {Ramolla}, M., {Pozo Nu{\~n}ez}, F., {Westhues}, C.,
  {Watermann}, R., {Hoffmeister}, V., \& {Murphy}, M. 2011, \aap, 535, A73

\bibitem[{{H{\"o}nig}(2014)}]{Honig2014}
{H{\"o}nig}, S.~F. 2014, \apjl, 784, L4

\bibitem[{{Hopkins} {et~al.}(2006){Hopkins}, {Hernquist}, {Cox}, {Di Matteo},
  {Robertson}, \& {Springel}}]{Hopkins2006}
{Hopkins}, P.~F., {Hernquist}, L., {Cox}, T.~J., {Di Matteo}, T., {Robertson},
  B., \& {Springel}, V. 2006, \apjs, 163, 1

\bibitem[{{Jiang} {et~al.}(2010){Jiang}, {Fan}, {Brandt}, {Carilli}, {Egami},
  {Hines}, {Kurk}, {Richards}, {Shen}, {Strauss}, {Vestergaard}, \&
  {Walter}}]{Jiang2010}
{Jiang}, L., {Fan}, X., {Brandt}, W.~N., {Carilli}, C.~L., {Egami}, E.,
  {Hines}, D.~C., {Kurk}, J.~D., {Richards}, G.~T., {Shen}, Y., {Strauss},
  M.~A., {Vestergaard}, M., \& {Walter}, F. 2010, \nat, 464, 380

\bibitem[{{Kaspi} {et~al.}(2007){Kaspi}, {Brandt}, {Maoz}, {Netzer},
  {Schneider}, \& {Shemmer}}]{Kaspi07}
{Kaspi}, S., {Brandt}, W.~N., {Maoz}, D., {Netzer}, H., {Schneider}, D.~P., \&
  {Shemmer}, O. 2007, \apj, 659, 997

\bibitem[{{Kaspi} {et~al.}(2005){Kaspi}, {Maoz}, {Netzer}, {Peterson},
  {Vestergaard}, \& {Jannuzi}}]{Kaspi2005}
{Kaspi}, S., {Maoz}, D., {Netzer}, H., {Peterson}, B.~M., {Vestergaard}, M., \&
  {Jannuzi}, B.~T. 2005, \apj, 629, 61

\bibitem[{{Kaspi} {et~al.}(2000){Kaspi}, {Smith}, {Netzer}, {Maoz}, {Jannuzi},
  \& {Giveon}}]{Kaspi2000}
{Kaspi}, S., {Smith}, P.~S., {Netzer}, H., {Maoz}, D., {Jannuzi}, B.~T., \&
  {Giveon}, U. 2000, \apj, 533, 631

\bibitem[{{Kaspi} {et~al.}(1996)}]{Kaspi96}
{Kaspi}, S., {et~al.} 1996, \apj, 470, 336

\bibitem[{{Kelly}(2007)}]{Kelly2007}
{Kelly}, B.~C. 2007, \apj, 665, 1489

\bibitem[{{Kelly} {et~al.}(2009){Kelly}, {Bechtold}, \&
  {Siemiginowska}}]{Kelly2009}
{Kelly}, B.~C., {Bechtold}, J., \& {Siemiginowska}, A. 2009, \apj, 698, 895

\bibitem[{{Kelly} \& {Shen}(2013)}]{KellyShen2013}
{Kelly}, B.~C., \& {Shen}, Y. 2013, \apj, 764, 45

\bibitem[{{Kelly} {et~al.}(2010){Kelly}, {Vestergaard}, {Fan}, {Hopkins},
  {Hernquist}, \& {Siemiginowska}}]{Kelly2010}
{Kelly}, B.~C., {Vestergaard}, M., {Fan}, X., {Hopkins}, P., {Hernquist}, L.,
  \& {Siemiginowska}, A. 2010, \apj, 719, 1315

\bibitem[{{King} {et~al.}(2014){King}, {Davis}, {Denney}, {Vestergaard}, \&
  {Watson}}]{King2014}
{King}, A.~L., {Davis}, T.~M., {Denney}, K.~D., {Vestergaard}, M., \& {Watson},
  D. 2014, \mnras, 441, 3454

\bibitem[{{Kinney} {et~al.}(1991){Kinney}, {Bohlin}, {Blades}, \&
  {York}}]{Kinney1991}
{Kinney}, A.~L., {Bohlin}, R.~C., {Blades}, J.~C., \& {York}, D.~G. 1991,
  \apjs, 75, 645

\bibitem[{{Koratkar} \& {Gaskell}(1991)}]{Koratkar1991}
{Koratkar}, A.~P., \& {Gaskell}, C.~M. 1991, \apjl, 370, L61

\bibitem[{{Korista} \& {Goad}(2004)}]{Korista2004}
{Korista}, K.~T., \& {Goad}, M.~R. 2004, \apj, 606, 749

\bibitem[{{Korista} {et~al.}(1995)}]{Korista95}
{Korista}, K.~T., {et~al.} 1995, \apjs, 97, 285

\bibitem[{{Koz{\l}owski} {et~al.}(2010)}]{Kozlowski2010}
{Koz{\l}owski}, S., {et~al.} 2010, \apj, 708, 927

\bibitem[{{Krogager} {et~al.}(2013)}]{Krogager2013}
{Krogager}, J.-K., {et~al.} 2013, \mnras, 433, 3091

\bibitem[{{Laor}(1998)}]{Laor1998}
{Laor}, A. 1998, \apjl, 505, L83

\bibitem[{{Lynden-Bell}(1969)}]{Lynden-Bell1969}
{Lynden-Bell}, D. 1969, \nat, 223, 690

\bibitem[{{MacLeod} {et~al.}(2010)}]{MacLeod2010}
{MacLeod}, C.~L., {et~al.} 2010, \apj, 721, 1014

\bibitem[{{Mathews} \& {Capriotti}(1985)}]{Mathews1984}
{Mathews}, W.~G., \& {Capriotti}, E.~R. 1985, in Astrophysics of Active
  Galaxies and Quasi-Stellar Objects, ed. J.~S. {Miller}, 185--233

\bibitem[{{McGill} {et~al.}(2008){McGill}, {Woo}, {Treu}, \&
  {Malkan}}]{McGill2008}
{McGill}, K.~L., {Woo}, J.-H., {Treu}, T., \& {Malkan}, M.~A. 2008, \apj, 673,
  703

\bibitem[{{McLure} \& {Jarvis}(2002)}]{McLureJarvis2002}
{McLure}, R.~J., \& {Jarvis}, M.~J. 2002, \mnras, 337, 109

\bibitem[{{McNamara} \& {Nulsen}(2007)}]{McNamaraNulsen2007}
{McNamara}, B.~R., \& {Nulsen}, P.~E.~J. 2007, \araa, 45, 117

\bibitem[{{Melia}(2014)}]{Melia2014}
{Melia}, F. 2014, J. Cosmology Astropart. Physis., 1, 27

\bibitem[{{Merloni} \& {Heinz}(2007)}]{Merloni2007}
{Merloni}, A., \& {Heinz}, S. 2007, \mnras, 381, 589

\bibitem[{{Meusinger} {et~al.}(2011){Meusinger}, {Hinze}, \& {de
  Hoon}}]{Meusinger2011}
{Meusinger}, H., {Hinze}, A., \& {de Hoon}, A. 2011, \aap, 525, A37

\bibitem[{{Mortlock} {et~al.}(2011)}]{Mortlock2011}
{Mortlock}, D.~J., {et~al.} 2011, \nat, 474, 616

\bibitem[{{O'Brien} {et~al.}(1998)}]{O'Brien98}
{O'Brien}, P.~T., {et~al.} 1998, \apj, 509, 163

\bibitem[{{Osterbrock} \& {Ferland}(2006)}]{Osterbrock2006}
{Osterbrock}, D.~E., \& {Ferland}, G.~J. 2006, {Astrophysics of gaseous nebulae
  and active galactic nuclei} (CA: University Science Books)

\bibitem[{{Paltani} \& {Courvoisier}(1994)}]{Paltani1994}
{Paltani}, S., \& {Courvoisier}, T.~J.-L. 1994, \aap, 291, 74

\bibitem[{{Park} {et~al.}(2013){Park}, {Woo}, {Denney}, \& {Shin}}]{Park2013}
{Park}, D., {Woo}, J.-H., {Denney}, K.~D., \& {Shin}, J. 2013, \apj, 770, 87

\bibitem[{{Peterson}(1988)}]{Peterson1988}
{Peterson}, B.~M. 1988, \pasp, 100, 18

\bibitem[{{Peterson}(1993)}]{Peterson93}
---. 1993, \pasp, 105, 247

\bibitem[{{Peterson}(1997)}]{Petersonbook}
---. 1997, {An Introduction to Active Galactic Nuclei} (Cambridge, New York
  Cambridge University Press)

\bibitem[{{Peterson}(2010)}]{Peterson2010}
{Peterson}, B.~M. 2010, in IAU Symposium, Vol. 267, IAU Symposium, 151--160

\bibitem[{{Peterson} {et~al.}(2013){Peterson}, {Denney}, {De Rosa}, {Grier},
  {Pogge}, {Bentz}, {Kochanek}, {Vestergaard}, {Kilerci-Eser}, {Dalla
  Bont{\`a}}, \& {Ciroi}}]{Peterson13}
{Peterson}, B.~M., {Denney}, K.~D., {De Rosa}, G., {Grier}, C.~J., {Pogge},
  R.~W., {Bentz}, M.~C., {Kochanek}, C.~S., {Vestergaard}, M., {Kilerci-Eser},
  E., {Dalla Bont{\`a}}, E., \& {Ciroi}, S. 2013, \apj, 779, 109

\bibitem[{{Peterson} {et~al.}(2002)}]{Peterson2002}
{Peterson}, B.~M., {et~al.} 2002, \apj, 581, 197

\bibitem[{{Peterson} {et~al.}(2004)}]{Peterson04}
---. 2004, \apj, 613, 682

\bibitem[{{Peterson} {et~al.}(2014)}]{Peterson2014}
---. 2014, ArXiv e-prints. arXiv:1409.4448

\bibitem[{{Press} {et~al.}(1992){Press}, {Teukolsky}, {Vetterling}, \&
  {Flannery}}]{Press92}
{Press}, W.~H., {Teukolsky}, S.~A., {Vetterling}, W.~T., \& {Flannery}, B.~P.
  1992, {Numerical recipes in FORTRAN. The art of scientific computing}
  (Cambridge: University Press)

\bibitem[{{Rafiee} \& {Hall}(2011)}]{Rafiee2011}
{Rafiee}, A., \& {Hall}, P.~B. 2011, \apjs, 194, 42

\bibitem[{{Rees}(1984)}]{Rees1984}
{Rees}, M.~J. 1984, \araa, 22, 471

\bibitem[{{Reichert} {et~al.}(1994)}]{Reichert94}
{Reichert}, G.~A., {et~al.} 1994, \apj, 425, 582

\bibitem[{{Richards} {et~al.}(2003)}]{Richards2003}
{Richards}, G.~T., {et~al.} 2003, \aj, 126, 1131

\bibitem[{{Rodriguez-Pascual} {et~al.}(1997)}]{Rodriguez-Pascual97}
{Rodriguez-Pascual}, P.~M., {et~al.} 1997, \apjs, 110, 9

\bibitem[{{Santos-Lleo} {et~al.}(1997)}]{Santos97}
{Santos-Lleo}, M., {et~al.} 1997, \apjs, 112, 271

\bibitem[{{Schlafly} \& {Finkbeiner}(2011)}]{Schlafly2011}
{Schlafly}, E.~F., \& {Finkbeiner}, D.~P. 2011, \apj, 737, 103

\bibitem[{{Schlegel} {et~al.}(1998){Schlegel}, {Finkbeiner}, \&
  {Davis}}]{Schlegel98}
{Schlegel}, D.~J., {Finkbeiner}, D.~P., \& {Davis}, M. 1998, \apj, 500, 525

\bibitem[{{Schmidt} {et~al.}(2010){Schmidt}, {Marshall}, {Rix}, {Jester},
  {Hennawi}, \& {Dobler}}]{Schmidt2010}
{Schmidt}, K.~B., {Marshall}, P.~J., {Rix}, H.-W., {Jester}, S., {Hennawi},
  J.~F., \& {Dobler}, G. 2010, \apj, 714, 1194

\bibitem[{{Sergeev} {et~al.}(2007){Sergeev}, {Doroshenko}, {Dzyuba},
  {Peterson}, {Pogge}, \& {Pronik}}]{sergeev07}
{Sergeev}, S.~G., {Doroshenko}, V.~T., {Dzyuba}, S.~A., {Peterson}, B.~M.,
  {Pogge}, R.~W., \& {Pronik}, V.~I. 2007, \apj, 668, 708

\bibitem[{{Shen} {et~al.}(2011){Shen}, {Richards}, {Strauss}, {Hall},
  {Schneider}, {Snedden}, {Bizyaev}, {Brewington}, {Malanushenko},
  {Malanushenko}, {Oravetz}, {Pan}, \& {Simmons}}]{Shen2011}
{Shen}, Y., {Richards}, G.~T., {Strauss}, M.~A., {Hall}, P.~B., {Schneider},
  D.~P., {Snedden}, S., {Bizyaev}, D., {Brewington}, H., {Malanushenko}, V.,
  {Malanushenko}, E., {Oravetz}, D., {Pan}, K., \& {Simmons}, A. 2011, \apjs,
  194, 45

\bibitem[{{Stirpe} {et~al.}(1994)}]{Stirpe94}
{Stirpe}, G.~M., {et~al.} 1994, \apj, 425, 609

\bibitem[{{Taylor}(1997)}]{Taylor}
{Taylor}, J. 1997, {Introduction to Error Analysis, the Study of Uncertainties
  in Physical Measurements, 2nd Edition} (University Science Books)

\bibitem[{{Trakhtenbrot} \& {Netzer}(2012)}]{Trakhtenbroot2012}
{Trakhtenbrot}, B., \& {Netzer}, H. 2012, \mnras, 427, 3081

\bibitem[{{Tremaine} {et~al.}(2002)}]{Tremaine2002}
{Tremaine}, S., {et~al.} 2002, \apj, 574, 740

\bibitem[{{Tully} {et~al.}(2009){Tully}, {Rizzi}, {Shaya}, {Courtois},
  {Makarov}, \& {Jacobs}}]{Tully2009}
{Tully}, R.~B., {Rizzi}, L., {Shaya}, E.~J., {Courtois}, H.~M., {Makarov},
  D.~I., \& {Jacobs}, B.~A. 2009, \aj, 138, 323

\bibitem[{{Vanden Berk} {et~al.}(2004)}]{Vandenberk2004}
{Vanden Berk}, D.~E., {et~al.} 2004, \apj, 601, 692

\bibitem[{{Vestergaard}(2002)}]{Vestergaard2002}
{Vestergaard}, M. 2002, \apj, 571, 733

\bibitem[{{Vestergaard}(2004)}]{Vestergaard2004}
---. 2004, \apj, 601, 676

\bibitem[{{Vestergaard} {et~al.}(2008){Vestergaard}, {Fan}, {Tremonti},
  {Osmer}, \& {Richards}}]{Vestergaard2008}
{Vestergaard}, M., {Fan}, X., {Tremonti}, C.~A., {Osmer}, P.~S., \& {Richards},
  G.~T. 2008, \apjl, 674, L1

\bibitem[{{Vestergaard} \& {Osmer}(2009)}]{Vestergaard2009}
{Vestergaard}, M., \& {Osmer}, P.~S. 2009, \apj, 699, 800

\bibitem[{{Vestergaard} \& {Peterson}(2006)}]{Vestergaard2006}
{Vestergaard}, M., \& {Peterson}, B.~M. 2006, \apj, 641, 689

\bibitem[{{Wandel} {et~al.}(1999){Wandel}, {Peterson}, \&
  {Malkan}}]{Wandel1999}
{Wandel}, A., {Peterson}, B.~M., \& {Malkan}, M.~A. 1999, \apj, 526, 579

\bibitem[{{Wanders} {et~al.}(1997)}]{Wanders97}
{Wanders}, I., {et~al.} 1997, \apjs, 113, 69

\bibitem[{{Wang} {et~al.}(2009)}]{Wang2009}
{Wang}, J.-G., {et~al.} 2009, \apj, 707, 1334

\bibitem[{{Warren} {et~al.}(1994){Warren}, {Hewett}, \& {Osmer}}]{WHO1994}
{Warren}, S.~J., {Hewett}, P.~C., \& {Osmer}, P.~S. 1994, \apj, 421, 412

\bibitem[{{Watson} {et~al.}(2011){Watson}, {Denney}, {Vestergaard}, \&
  {Davis}}]{Watson2011}
{Watson}, D., {Denney}, K.~D., {Vestergaard}, M., \& {Davis}, T.~M. 2011,
  \apjl, 740, L49

\bibitem[{{Werner} {et~al.}(2014)}]{Werner2013}
{Werner}, N., {et~al.} 2014, \mnras, 439, 2291

\bibitem[{{Wilhite} {et~al.}(2005)}]{Wilhite2005}
{Wilhite}, B.~C., {et~al.} 2005, \apj, 633, 638

\bibitem[{{Wolfe} {et~al.}(2005){Wolfe}, {Gawiser}, \& {Prochaska}}]{Wolfe2005}
{Wolfe}, A.~M., {Gawiser}, E., \& {Prochaska}, J.~X. 2005, \araa, 43, 861

\bibitem[{{Wu} {et~al.}(2004){Wu}, {Wang}, {Kong}, {Liu}, \& {Han}}]{Wu2004}
{Wu}, X.-B., {Wang}, R., {Kong}, M.~Z., {Liu}, F.~K., \& {Han}, J.~L. 2004,
  \aap, 424, 793

\bibitem[{{Yoshii} {et~al.}(2014){Yoshii}, {Kobayashi}, {Minezaki}, {Koshida},
  \& {Peterson}}]{Yoshii2014}
{Yoshii}, Y., {Kobayashi}, Y., {Minezaki}, T., {Koshida}, S., \& {Peterson},
  B.~A. 2014, \apjl, 784, L11

\bibitem[{{Zu} {et~al.}(2011){Zu}, {Kochanek}, \& {Peterson}}]{Zu2011}
{Zu}, Y., {Kochanek}, C.~S., \& {Peterson}, B.~M. 2011, \apj, 735, 80

\bibitem[{{Zuo} {et~al.}(2012){Zuo}, {Wu}, {Liu}, \& {Jiao}}]{Zuo2012}
{Zuo}, W., {Wu}, X.-B., {Liu}, Y.-Q., \& {Jiao}, C.-L. 2012, \apj, 758, 104

\end{thebibliography}

\end{document}